\documentclass[a4paper,fleqn,number]{cas-sc}

\usepackage[numbers]{natbib}
\usepackage{subcaption}
\usepackage{indentfirst}
\usepackage{microtype}
\usepackage{siunitx}
\usepackage{placeins}
\usepackage{setspace}
\usepackage{lineno}

\def\tsc#1{\csdef{#1}{\textsc{\lowercase{#1}}\xspace}}
\tsc{WGM}
\tsc{QE}
\tsc{EP}
\tsc{PMS}
\tsc{BEC}
\tsc{DE}

\makeatletter
\renewcommand\subsection{\@startsection
  {subsection}{2}{\z@}%
  {-3.25ex\@plus -1ex \@minus -.2ex}%
  {1.5ex \@plus .2ex}%
  {\normalfont\normalsize\itshape}}
\makeatother

\begin{document}
\modulolinenumbers[1]

\let\WriteBookmarks\relax
\def\floatpagepagefraction{1}
\def\textpagefraction{.001}
\shorttitle{Inertial effects on the mechanical efficiency of a foil-based energy harvester}
\shortauthors{Z.Zihan et~al.}

\title [mode = title]{Inertial effects on the mechanical efficiency of a semi-passive oscillating hydrofoil energy harvester}                      

\tnotetext[1]{This project is funded by Iowa Energy Center. YZ acknowledges support from the UC Riverside Winston Chung Global Energy Center and the SoCal OASIS Internal Funding Award.}

\author[1]{Zihan Zhang}[type=editor,
                        auid=000,bioid=1,
                        orcid=0009-0002-2797-9549]
\cormark[1]
\ead{zihan34@iastate.edu}

\credit{Conceptualization of this study, Methodology, Software}

\affiliation[1]{organization={Department of Mechanical Engineering},
                addressline={Iowa State University}, 
                city={Ames},
                postcode={50011}, 
                state={IA},
                country={United States}}

\author[1]{Qimin Feng}[style=chinese, orcid=0009-0002-1035-7313]
\ead{qmfeng11@iastate.edu}

\credit{Methodology--PIV Experiments}

\author[2]{Yuanhang Zhu}[style=chinese, orcid=0000-0002-2080-1142]

\ead{yuanhang.zhu@ucr.edu}

\credit{Writing - Original draft preparation}

\affiliation[2]{organization={Department of Mechanical Engineering},
                addressline={University of California}, 
                city={Riverside},
                postcode={92521}, 
                state={CA},
                country={United States}}

\author[1]{Qiang Zhong}[style=chinese, orcid=0000-0002-8435-5938]
\ead{qzhong1@iastate.edu}
\credit{Data curation, Writing - Original draft preparation}

\cortext[cor1]{Corresponding author}

\begin{abstract}
Oscillating-foil-based energy harvesters have demonstrated strong potential for low-speed hydrokinetic energy extraction; however, the actuator-level mechanical energy balance associated with prescribed pitching motion remains poorly understood. The present work experimentally characterizes how foil mass ratio, pitching-axis location, and reduced frequency jointly govern the hydrodynamic and mechanical efficiencies of a semi-passive oscillating hydrofoil. Results show that rotational inertia redistributes actuator demand through phase-dependent torque exchange, while heave--pitch coupling can partially cancel this demand when favorably phased. Pitching-axis location modifies the phase and direction of the fluid torque through changes in the effective hydrodynamic moment arm. Reduced frequency governs the balance between enhanced unsteady loading and inertia-amplified actuator demand. Optimal performance is achieved within reduced frequency region of 0.125--0.16 using quarter-chord to one-third-chord pitching axes and relatively low foil mass ratios from about 0.5 to 2.0, yielding a peak mechanical efficiency of 33.96\%---which can diverge from the hydrodynamic efficiency by approximately 38.16\% depending on configuration. Torque-loop analysis and PIV measurements show that this synchronization is a key mechanism governing the observed efficiency trends.

\end{abstract}

\begin{graphicalabstract}
\begin{figure*}
    \centering
    \includegraphics[width=1.0\linewidth]{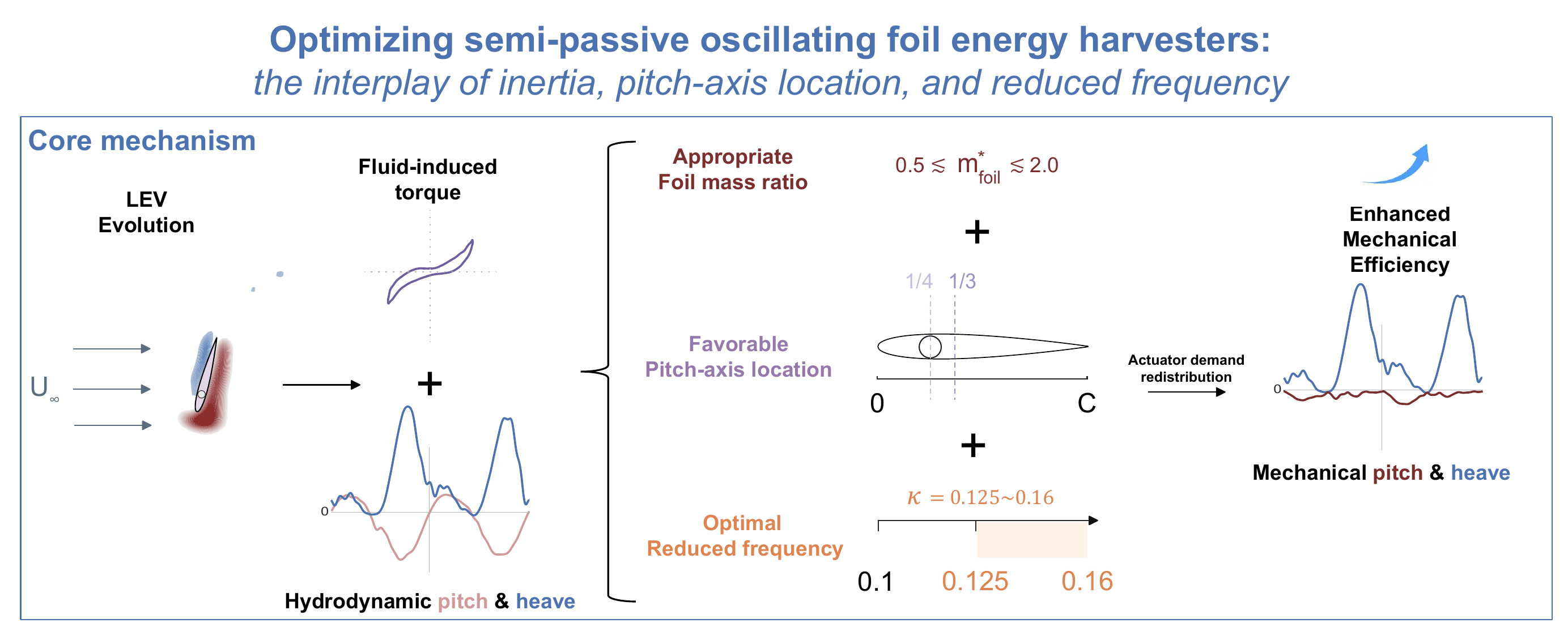}
    \label{fig:graphic abstract}
\end{figure*}

\end{graphicalabstract}

\begin{highlights}
\item Inertia redistributes actuator demand through phase-dependent torque exchange.
\item Pitch axis controls fluid-torque direction through moment-arm variation.
\item Reduced frequency amplifies both fluid forcing and inertial loading.
\end{highlights}


\maketitle
\doublespacing

\section{Introduction}
Hydrokinetic energy from tidal and river currents represents a predictable and widely distributed renewable-energy resource. Oscillating hydrofoils have emerged as a promising alternative to conventional rotary turbines for harvesting this resource, offering competitive efficiency together with distinct advantages in low-speed and shallow-water environments, reduced susceptibility to debris, and favorable wake characteristics for array deployment \cite{platzer2009development,handy2025optimal,ribeiro2021wake}. Unlike rotating blades, oscillating foils extract energy through coupled heaving and pitching motions, leveraging unsteady hydrodynamic forces to convert fluid kinetic energy into mechanical power \cite{jones2003investigation,kinsey2011prototype,mckinney1981wingmill}.

Over the past two decades, extensive research~\cite{xiao2014review} has mapped the performance of oscillating hydrofoil energy harvesters across three main system configurations. In \textit{fully active systems}, where both heaving and pitching motions are prescribed, parametric optimization has yielded peak hydrodynamic efficiencies of $30$--$35\%$ at a reduced frequency of $0.15$~\cite{ribeiro2020vortex}, a pitch amplitude of $75^\circ$~\cite{kinsey2008parametric}, and a $\pi/2$ phase difference between heave and pitch \cite{simpson2008experiments}. In \textit{fully passive systems}, both degrees of freedom are driven by hydrodynamically induced oscillations through tuned mechanical parameters~\cite{duarte2019experimental}, with reported efficiencies up to $41\%$ in simulations~\cite{young2013numerical} and around $31$--$43\%$ in experiments \cite{boudreau2018experimental,duarte2021experimental,zhao2021flow}. \textit{Semi-passive systems}, in which one degree of freedom---typically pitch---is prescribed while the other responds freely to fluid forcing, combine controllability with flow-adaptive dynamics and have demonstrated efficiencies exceeding $30\%$ ~\cite{deng2015inertial,teng2016effects}. Beyond these baseline configurations, performance enhancements have also been pursued through effective angle of attack~\cite{boudis2021effects}, non-sinusoidal pitching profiles ~\cite{lu2014nonsinusoidal,teng2016effects,xiao2012motion}, wall confinement effects ~\cite{karakas2016effect,su2019confinement}, aspect-ratio variations ~\cite{deng2014effect,kim2017energy}, and tandem arrangements that exploit wake--foil vortex interactions ~\cite{ashraf2011numerical,ribeiro2024prediction,simeski2017simulations}.

Despite these advances, most reported efficiencies are hydrodynamic in nature, computed from the cycle-averaged energy exchange between the fluid and the foil~\cite{kinsey2008parametric,xiao2014review,zhu2011optimal}. This metric is essential for comparing foil kinematics, geometries, and operating conditions, but it does not fully represent the net power balance of a semi-passive energy-harvesting device. In particular, hydrodynamic efficiency is mathematically insensitive to foil inertia: although the foil has finite rotational inertia, the corresponding inertial power integrates to zero over a periodic cycle and therefore does not contribute to the cycle-averaged numerator. Thus, hydrodynamic efficiency characterizes the fluid--foil energy exchange, independent of how the pitching motion is actuated.

This cancellation does not carry over to the engineering power consumption of a real semi-passive energy harvester, where only the pitching motion is actively prescribed while the heaving motion responds passively to fluid loading. In a non-regenerative pitch drive---as used in many practical hydrofoil rigs~\cite{kinsey2011prototype,su2019resonant}---the work required to accelerate the foil against inertia cannot be recovered during deceleration; instead, it is dissipated through active braking, motor losses, or transmission losses. As a result, the cancellation embedded in the hydrodynamic cycle average is replaced by an accumulation of instantaneous actuator-power demand. At the same time, heave--pitch coupling through the foil static moment introduces an additional energy pathway that is absent from the hydrodynamic-efficiency framework~\cite{boudreau2018experimental,veilleux2017numerical}. Depending on the relative position of the pitching axis and the center of mass, this coupling can either alleviate or exacerbate the mechanical power demand. This distinction makes \textit{mechanical efficiency}, rather than \textit{hydrodynamic efficiency} alone, the relevant metric for evaluating actuator-level performance in practical semi-passive hydrofoil energy harvesters.

A limited number of studies have begun to address the role of structural inertia in governing the efficiency of oscillating-foil energy harvesting. Deng et al.~\cite{deng2015inertial} examined how mass ratio modulates energy extraction in semi-passive systems, finding that total power extraction remains nearly unchanged for moderate mass ratios, whereas the hydrodynamic energy-harvesting efficiency decreases monotonically with increasing mass ratio. However, their analysis did not address the actuator-side mechanical power cost associated with non-regenerative prescribed pitching motion. Mackowski and Williamson~\cite{mackowski2017effect}, Wang et al.~\cite{wang2021pivot}, Zhu et al.~\cite{zhu2021nonlinear}, and Zhu et al.~\cite{zhu2021improve} showed that pivot location and mass ratio strongly influence the dynamic response of passively heaving and fully passive foils. More recently, Siala et al.~\cite{siala2020experimental} demonstrated experimentally that inertia-driven mechanisms can enhance energy extraction in a phase-dependent manner. Nevertheless, a systematic investigation of how inertia influences overall \textit{mechanical efficiency}, particularly across the combined parameter space of mass ratio, pitching-axis location, and reduced frequency, remains lacking. This gap directly limits the translation from laboratory hydrodynamic characterization to engineering system design, since these are precisely the parameters that must be specified in a field-deployable hydrokinetic device.

Motivated by this gap, the present study experimentally investigates how foil mass ratio, pitching-axis location, and reduced frequency jointly shape the mechanical efficiency of a semi-passive oscillating hydrofoil energy harvester using a cyber-physical experimental system. These three parameters affect the actuator-level energy balance in distinct but coupled ways: mass ratio scales inertial loading, pitching-axis location changes the static moment and hydrodynamic moment arm, and reduced frequency amplifies inertial torque while modulating its phase relative to fluid forcing. We show that mechanical efficiency can deviate substantially from hydrodynamic efficiency: inertia can reduce pitch-power consumption for upstream-of-center-of-mass pivots, but can increase actuator demand when the pivot is near or downstream of the center of mass. Within the tested parameter space, the highest mechanical efficiency occurs near the one-third-chord pivot, whereas large mass ratios at mid-chord produce a severe mechanical-efficiency penalty. These results identify an operating window in which structural inertia improves device-level energy conversion and demonstrate why mechanical efficiency should be considered alongside hydrodynamic efficiency in semi-passive oscillating-foil energy harvesters.

\section{Methods}

\begin{figure}
    \centering
    
    \begin{subfigure}{0.9\textwidth}
        \centering
        \includegraphics[width=\linewidth]{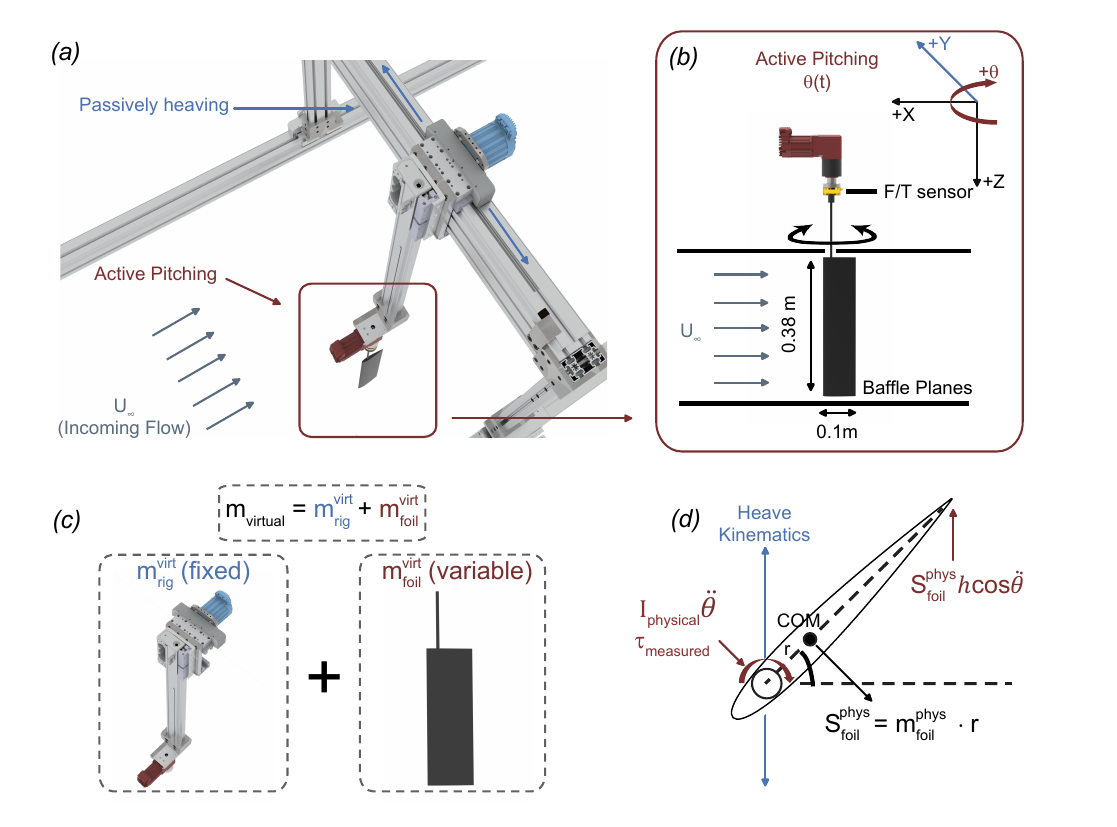}
    \end{subfigure}

    \caption{(a) Experimental platform: Schematic of the hydrofoil installed in the water channel. (b) Operating principle and measurements: Side-view diagram of the pitching hydrofoil, force transducer, and baffle planes. (c) Mass decomposition: total mass ratio: $m_\mathrm{virtual}^* =  m_\mathrm{rig}^* + m_\mathrm{foil}^* =(m^\mathrm{virt}_\mathrm{rig} + m^\mathrm{virt}_\mathrm{foil}) / 0.5\rho c^2 s$; foil mass ratio: $m_\mathrm{foil}^* = m^\mathrm{virt}_\mathrm{foil} / 0.5\rho c^2 s$. (d) Torque decomposition: Top-down diagram of the hydrofoil. The unfilled circle denotes the pitch-axis location, and $r$ is the distance between the center of mass (COM) and the pivot point.}
    \label{Fig:1}
\end{figure}

\subsection{\textit{Experimental setup}}

As shown in Fig.~\ref{Fig:1}, experiments were conducted in a closed-loop water channel with a test section measuring $3\,\mathrm{m} \times 1\,\mathrm{m} \times 0.5\,\mathrm{m}$ (length $\times$ width $\times$ height). A NACA0012 hydrofoil with chord length $c = 0.1\,\mathrm{m}$, span $s = 0.38\,\mathrm{m}$, and aspect ratio $AR = s/c = 3.8$ was mounted vertically on a 6-axis force/torque transducer (ATI Mini40 IP65). The hydrofoil was 3D-printed from ABS using a Bambu X1 printer and actuated through a $7\,\mathrm{mm}$-diameter carbon-fiber shaft installed along its pitching axis. Another carbon-fiber tube with a $10\,\mathrm{mm}$ outer diameter was used as a reinforcement sleeve to reduce shaft deflection under fluid loading.

Four hydrofoils with identical mass were printed to accommodate the four pitching-axis locations investigated in this study: $x/c = 0$, $0.25$, $0.33$, and $0.5$. The measured physical mass of each foil was approximately $m^{\mathrm{phys}}_{\mathrm{foil}} \approx 0.29\,\mathrm{kg}$. Acrylic baffles were positioned $5\,\mathrm{mm}$ above the foil tip and along the bottom of the water channel to approximate two-dimensional flow conditions.

The pitching motion was driven by a Teknic servomotor (CPM-MCPV-2341S-ELS), whereas the heaving motion, perpendicular to the freestream, was driven by a second servomotor (CPM-MCPV-3432P-ELS) operating under cyber-physical control (Section~2.2). For all experiments, the freestream velocity was maintained at $U_{\infty} = 0.4\,\mathrm{m/s}$ and validated using an ultrasound flow sensor (TUF-2000B), corresponding to a chord-based Reynolds number of $Re = \rho U_{\infty} c / \mu \approx 4 \times 10^4$. The pitch position $\theta(t)$ was measured in real time using an optical encoder (US Digital, 5000\,CPR, EM2-2-5000-I) attached to the pitch motor. Forces and torques were sampled at $4000\,\mathrm{Hz}$ using the force/torque transducer and a 16-bit analog-to-digital converter (NI PCIe-6363). All data acquisition and motor control were managed through a Simulink-based program running on the control PC.

\subsection{\textit{Cyber-physical system (CPS)}}

In this study, the pitching motion was prescribed, whereas the heaving degree of freedom was realized through a cyber-physical system that emulated a virtual spring--mass--damper mechanism in real time. Rather than relying on physical springs and dampers, the CPS drove the heave dynamics entirely through software, allowing the virtual mass, damping, and stiffness to be controlled independently without modifying the physical hardware. This approach, originally introduced to fluid dynamics experiments by Mackowski and Williamson~\cite{mackowski2011developing} and subsequently adopted in energy-harvesting studies~\cite{onoue2015large,su2019resonant,zhu2020nonlinear}, provides the flexibility needed to systematically vary the mass ratio while maintaining identical hydrodynamic conditions.

For a prescribed pitching and passive heaving foil considered in this study, the governing equation for the heave motion is:
\begin{equation}
  m_\mathrm{total}\ddot{h} + c_\mathrm{d}\dot{h} + kh 
  + S_\mathrm{foil}(\ddot{\theta}\cos\theta - \dot{\theta}^2\sin\theta) 
  = L_\mathrm{fluid},
  \label{eq:passive heave}
\end{equation}
where $m_\mathrm{total}$, $c_\mathrm{d}$, and $k$ are the effective total mass, damping, and stiffness of the heaving axis, respectively. The effective total mass is defined as $m_\mathrm{total} = m_\mathrm{physical} + m_\mathrm{virtual}$, where $m_\mathrm{physical}$ and $m_\mathrm{virtual}$ are the combination of the rig and foil, which are expressed as:
\begin{equation}
  m_\mathrm{physical} = m^\mathrm{phys}_\mathrm{rig} + m^\mathrm{phys}_\mathrm{foil},
  \label{eq:rig mass definition}
\end{equation}
and
\begin{equation}
  m_\mathrm{virtual} = m^\mathrm{virt}_\mathrm{rig} + m^\mathrm{virt}_\mathrm{foil}.
  \label{eq:foil mass definition}
\end{equation}
The heave displacement is denoted by $h$, and $L_\mathrm{fluid}$ is the lift force measured by the force/torque transducer after transformation into the lift direction. The effective static moment is $S_\mathrm{foil} = m_\mathrm{foil} r$, where $m_\mathrm{foil} = m^\mathrm{phys}_\mathrm{foil} + m^\mathrm{virt}_\mathrm{foil}$ , $r$ is the distance between the foil center of mass and the pitching axis~\cite{boudreau2020parametric,oshkai2022reliability}. The inertial coupling term $S_\mathrm{foil}(\ddot{\theta}\cos\theta - \dot{\theta}^2\sin\theta)$ accounts for the inertial interaction between the prescribed pitching motion and the passive heave response. Oshkai et al.~\cite{oshkai2022reliability} explained that this inertial coupling term existed due to the static imbalance $r$ in Fig.~\ref{Fig:1}(d).

The CPS operated in a position-command mode. At each time step, the lift force $L_\mathrm{fluid}$ was calculated from the real-time force measurements, $F_{\mathrm{x}}(t)$, $F_{\mathrm{y}}(t)$, and the instantaneous pitch angle $\theta(t)$ measured by the encoder based on our current system setup:
\begin{equation}
  L_\mathrm{fluid}(t) = F_{\mathrm{x}}(t)\sin\theta(t) + F_{\mathrm{y}}(t)\cos\theta(t).
  \label{eq:lift force}
\end{equation}
According to Eq.~\ref{eq:passive heave}, the present system does not include a physical spring mechanism, and the intrinsic mechanical damping of the setup is sufficiently small to be neglected. Therefore, the target stiffness and damping characteristics were imposed through virtual control parameters, such that $k = k_\mathrm{virtual}$ and $c_\mathrm{d} = c_\mathrm{d,virtual}$. By combining the virtual inertial, damping, stiffness, and coupling contributions, the equivalent structural force applied to the passive heave motion can be expressed as:
\begin{equation}
  L_\mathrm{structure} = -(m_\mathrm{virtual}\ddot{h} + c_\mathrm{d}\dot{h} + kh + S^\mathrm{virt}_\mathrm{foil}(\ddot{\theta}\cos\theta - \dot{\theta}^2\sin\theta)).
  \label{eq:structural lift}
\end{equation}
Substituting equation~\ref{eq:structural lift} into~\ref{eq:passive heave}, yields the governing equation for the passive heave acceleration:
\begin{equation}
  m_\mathrm{physical}\ddot{h} 
  + S^\mathrm{phys}_\mathrm{foil}(\ddot{\theta}\cos\theta - \dot{\theta}^2\sin\theta) 
  = L_\mathrm{fluid} + L_\mathrm{structure},
  \label{eq:heave acceleration}
\end{equation}
since $S^\mathrm{phys}_\mathrm{foil} = S_\mathrm{foil} - S^\mathrm{virt}_\mathrm{foil} = m_\mathrm{foil} r - m^{virt}_\mathrm{foil} r = m^{phys}_\mathrm{foil} r$.

Therefore, the heave acceleration $\ddot{h}$ was then computed from Eq.~\ref{eq:heave acceleration}, and the corresponding velocity $\dot{h}$ and displacement $h$ were obtained through numerical integration. The updated heave displacement was then sent to the heave motor as the next position command. This closed-loop cycle ran at a sampling rate of $4000\,\mathrm{Hz}$, ensuring that the virtual dynamics were enforced with sufficient temporal resolution. Although position-command methods generally have slower transient responses than force-command alternatives~\cite{lee2011virtual,mackowski2011developing}, they offer high stability and allow the virtual mass to differ substantially from the physical foil mass~\cite{mackowski2017effect}. This capability is essential for the present parametric study.

Following the non-dimensionalization of Su and Breuer~\cite{su2019resonant}, all system parameters were scaled using the characteristic hydrodynamic force $0.5\rho U_\infty^2 c s$:
\begin{equation*}
  m^* = \frac{m}{0.5\rho c^2 s}, ~
  c_d^* = \frac{c_d}{0.5\rho U_\infty c s}, ~
  k^* = \frac{k}{0.5\rho U_\infty^2 s}.
\end{equation*}

\subsection{Power and efficiency definitions}

The energy-harvesting performance of the system was characterized by the heave power extracted from the flow and the pitch power required to sustain the prescribed pitching motion. All power quantities are reported as dimensionless coefficients normalized by $0.5\rho U_\infty^3 c s$, whereas efficiencies are normalized by the available flow power $0.5\rho U_\infty^3 A_s$, where $A_s$ is the foil swept area, defined as the product of the peak-to-peak heave amplitude, $2h$, and the foil span, $s$~\cite{su2019resonant,su2019confinement}. Instantaneous quantities are written with an explicit time argument, e.g.\ $C_h(t)$, and their cycle averages are denoted by an overbar, $\bar{C}_h = \langle C_h(t)\rangle$, where $\langle\cdot\rangle$ denotes averaging over one oscillation period.

The torque measured by the force/torque transducer, $\tau_{\mathrm{measured}}(t)$, represents the net torque transmitted through the pitching shaft and includes contributions from hydrodynamic loading, the physical inertia and the heave-induced effect of the lightweight ABS foil. It is therefore interpreted as the fluid-induced torque about the pitching axis:
\begin{equation}
  \tau_{\theta}^{\,\mathrm{fluid}}(t) = \tau_{\mathrm{measured}}(t).
  \label{eq:fluid torque}
\end{equation}
The virtual inertial torque, $\tau_{\theta}^{\,\mathrm{inertial}}(t)$, was computed by the virtual moment of inertia, $I_{\mathrm{virtual}}$, which was obtained from the prescribed virtual foil mass, $m^\mathrm{virt}_{\mathrm{foil}} = m^{*}_{\mathrm{foil}}\,0.5\rho c^2 s$, using the same geometry:
\begin{equation}
  I_{\mathrm{virtual}} =
  \frac{m^\mathrm{virt}_{\mathrm{foil}}}{m^{\mathrm{phys}}_{\mathrm{foil}}}
  I_{\mathrm{physical}},
  \label{eq:virtual_inertia}
\end{equation}
the virtual inertial torque was then expressed as:
\begin{equation}
\tau_{\theta}^{\,\mathrm{inertial}}(t) = I_{\mathrm{virtual}}\ddot{\theta}(t).
\label{eq:inertial torque}
\end{equation}
The virtual heave-induced torque, $\tau_\mathrm{heave}(t)$, was computed by the virtual foil mass, $m^\mathrm{virt}_{\mathrm{foil}}$, with $S^\mathrm{virt}_{\mathrm{foil}} = m^{\mathrm{virt}}_{\mathrm{foil}} r$, so that all inertial and coupling terms consistently reflected the prescribed mass ratio~\cite{boudreau2018experimental,sorensen2024developing}:
\begin{equation}
\tau_{\mathrm{heave}}(t) = S^\mathrm{virt}_\mathrm{foil}\ddot{h}(t)\cos\theta(t).
\label{eq:heave-induced torque}
\end{equation}
Taking together, to evaluate the mechanical torque, $\tau_{\theta}^{\,\mathrm{mech}}(t)$, associated with a foil of arbitrary virtual mass, two corrections were applied:
\begin{equation}
\begin{aligned}
  \tau_{\theta}^{\,\mathrm{mech}}(t) =
  \tau_{\theta}^{\,\mathrm{fluid}}(t)
  + \tau_{\theta}^{\,\mathrm{inertial}}(t)
  + \tau_{\mathrm{heave}}(t),
  \label{eq:corrected torque}
\end{aligned}
\end{equation}
also as:
\begin{equation}
\begin{aligned}
  \tau_{\theta}^{\,\mathrm{mech}}(t) =
  \tau_{\mathrm{measured}}(t)
  + I_{\mathrm{virtual}}\ddot{\theta}(t)
  + S^\mathrm{virt}_\mathrm{foil}\ddot{h}(t)\cos\theta(t).
  \label{eq:corrected_torque}
\end{aligned}
\end{equation}
In Equation.~\ref{eq:corrected_torque}, the first correction adds the corresponding inertial torque, $\tau_{\theta}^{\,\mathrm{inertial}}(t)$ (Eq.~\ref{eq:inertial torque}) for the virtual foil mass, and the second accounts for the translational--rotational coupling torque, $\tau_{\mathrm{heave}}(t)$ (Eq.~\ref{eq:heave-induced torque}) from the virtual foil mass that is absent from the sensor measurement. 




The prescribed pitching motion was:
\begin{equation}
  \theta(t) = \theta_0\cos(2\pi f t),
  \label{eq:pitch_motion}
\end{equation}
with angular velocity and acceleration given by:
\begin{align}
  \dot{\theta}(t) &= -\theta_0(2\pi f)\sin(2\pi f t), \label{eq:pitch_velocity}\\
  \ddot{\theta}(t) &= -\theta_0(2\pi f)^2\cos(2\pi f t). \label{eq:pitch_acceleration}
\end{align}

Three instantaneous power coefficients were defined directly from the time-resolved force and torque signals:
\begin{equation}
\begin{aligned}
C_h(t) 
&= \frac{L_\mathrm{fluid}(t)\dot{h}(t)}{0.5\rho U_\infty^3 c s}, \\[5pt]
C_{P,\theta}^{\mathrm{hydro}}(t) 
&= \frac{\tau_{\theta}^{\,\mathrm{fluid}}(t)\dot{\theta}(t)}
{0.5\rho U_\infty^3 c s}, \\[5pt]
C_{P,\theta}^{\mathrm{mech}}(t) 
&= -\frac{\left|\tau_{\theta}^{\,\mathrm{mech}}(t)\dot{\theta}(t)\right|}
{0.5\rho U_\infty^3 c s}.
\end{aligned}
\label{eq:power_coefficients}
\end{equation}

Here, $L_\mathrm{fluid}(t)$ is the lift force entering the heave equation, and $\dot{h}(t)$ is the heave velocity. The heave power coefficient $C_h(t)$ is positive on average and represents the useful power output. The hydrodynamic pitch-power coefficient $C_{P,\theta}^{\mathrm{hydro}}(t)$ uses the corrected fluid torque, $\tau_\theta^{\mathrm{fluid}}(t)$. The mechanical pitch-power coefficient $C_{P,\theta}^{\mathrm{mech}}(t)$ uses a different convention: the absolute value $|\tau_{\theta}^{\,\mathrm{mech}}(t)\dot{\theta}(t)|$ captures the actuator's instantaneous electrical demand in the non-regenerative regime, where both driving and braking phases dissipate energy since no generator is connected to recover energy from the pitch motion, and the negative sign maintains the convention that pitch power is an energy cost.

The cycle-averaged power coefficients are:
\begin{equation}
\begin{aligned}
\bar{C}_h 
&= \left\langle C_h(t) \right\rangle, \\[4pt]
\bar{C}_{P,\theta}^{\mathrm{hydro}} 
&= \left\langle C_{P,\theta}^{\mathrm{hydro}}(t) \right\rangle, \\[4pt]
\bar{C}_{P,\theta}^{\mathrm{mech}} 
&= \left\langle C_{P,\theta}^{\mathrm{mech}}(t) \right\rangle .
\end{aligned}
\label{eq:cycle_averaged_power}
\end{equation}
The hydrodynamic and mechanical efficiencies are then defined as:
\begin{equation}
  \varepsilon_{\mathrm{hydro}} = \frac{c\,s}{A_s}\Bigl(\bar{C}_h + \bar{C}_{P,\theta}^{\mathrm{hydro}}\Bigr),
  \label{eq:hydro_efficiency}
\end{equation}
and
\begin{equation}
  \varepsilon_{\mathrm{mech}} = \frac{c\,s}{A_s}\Bigl(\bar{C}_h + \bar{C}_{P,\theta}^{\mathrm{mech}}\Bigr),
  \label{eq:mech_efficiency}
\end{equation}

The difference $\varepsilon_{\mathrm{hydro}} - \varepsilon_{\mathrm{mech}}$ quantifies the efficiency penalty imposed by structural inertia and heave--pitch coupling under non-regenerative actuator constraints, and constitutes the central quantity of interest in this study. These three power coefficients and two efficiency metrics are used consistently in Figs.~\ref{FIG:2}--\ref{FIG:6} and Sections~3.1--3.3.

\subsection{Experimental procedure}

\begin{table}[width=1.0\textwidth,cols=5,pos=h]
\caption{Experimental parameters.}
\label{tbl1}
\begin{tabular*}{\tblwidth}{@{} CCCCC @{} }
\toprule
$x/c$ & $m_\mathrm{foil}^*$ & $m_\mathrm{rig}^*$ & $k^*$ & $\kappa \quad(=f c/U_{\infty})$\\
\midrule

\multirow{7}{*}{[0,\, 0.25,\, 0.33,\, 0.5]}
& 0.5 & \multirow{7}{*}{1.9} & [0.95,\, 1.48,\, 1.96,\, 2.43] & \multirow{7}{*}{[0.1,\, 0.125,\, 0.144,\, 0.16]} \\
& 0.9 &                      & [1.14,\, 1.71,\, 2.29,\, 2.86] &                       \\
& 1.3 &                      & [1.26,\, 1.97,\, 2.62,\, 3.23] &                       \\
& 1.7 &                      & [1.42,\, 2.22,\, 2.95,\, 3.64] &                       \\
& 2.1 &                      & [1.58,\, 2.47,\, 3.27,\, 4.04] &                       \\
& 2.5 &                      & [1.74,\, 2.71,\, 3.60,\, 4.45] &                       \\
& 3.0 &                      & [2.00,\, 3.00,\, 4.00,\, 5.00] &                       \\

\bottomrule
\end{tabular*}
\end{table}

The pitching motion was prescribed as $\theta(t)=\theta_0\cos(2\pi f t)$ with a fixed amplitude of $\theta_0=75^\circ \approx 1.31\,\mathrm{rad}$, corresponding to optimal energy-harvesting conditions reported in prior studies~\cite{deng2015inertial,kinsey2008parametric}. The heaving response was governed by the CPS described in Section~2.2, with the virtual damping coefficient fixed at $c_d^*=1.5$ across all tests. A full-factorial design was employed over foil mass ratio, reduced frequency, and pitching-axis location, yielding 112 unique test conditions. The foil mass ratio and heaving dynamics were controlled virtually through the CPS, the reduced frequency was varied by changing the imposed pitching frequency, and the pitching-axis location was varied by repositioning the shaft. The full parameter space and fixed conditions are summarized in Table~\ref{tbl1}.

For each virtual mass ratio, $m_\mathrm{virtual}^*=m_\mathrm{rig}^*+m_\mathrm{foil}^*$, the stiffness coefficient $k^*$ was adjusted so that the natural frequency of the virtual heaving system, $f_n=\sqrt{k/m}/2\pi$, matched the prescribed pitching frequency. This imposed the resonance condition $f/f_n=1$ for all tested reduced frequencies. The corresponding stiffness values are listed in Table~\ref{tbl1}.

For each condition, the system was allowed to reach a statistically periodic state over the first 10 cycles, after which data were recorded for 30 consecutive cycles. Phase-averaged quantities were computed over these recorded cycles, and the reported efficiencies and power coefficients represent cycle-averaged values. Measurement uncertainty was estimated from cycle-to-cycle variability and is indicated by error bars in the figures.

\subsection{\textit{Particle image velocimetry}}

Two-dimensional particle image velocimetry (PIV) was used to characterize the phase-resolved vortex dynamics associated with the oscillating hydrofoil. The flow was seeded with \SI{60}{\micro\meter} polyamide particles and illuminated by a 10~W laser sheet system (LWPIV-XH-10). Images with a resolution of $4096 \times 1532$ pixels were acquired using a high-speed camera (Photron Nova R3-4K) synchronized with the prescribed pitching motion. Three PIV conditions were tested to isolate the effect of pitching-axis location on the fluid--structure interaction (Table~\ref{tbl2}).

Phase-averaged vorticity fields were constructed from 312 instantaneous frames over three oscillation cycles, yielding 104 phase positions per cycle for each configuration. The velocity fields were processed in LaVision DaVis 10 using a standard cross-correlation algorithm with overlapping $128 \times 128$ pixel interrogation windows. The resulting phase-resolved vorticity fields were used to relate LEV evolution to the measured torque and power characteristics.

\begin{table}[width=.48\textwidth,cols=5,pos=h]
\caption{PIV experimental parameters.}
\label{tbl2}
\begin{tabular*}{\tblwidth}{@{} CCC @{} }
\toprule
 & $x/c$ & $\kappa \quad(=f c/U_{\infty})$\\
\midrule

Pitching-axis location effect & [0,\, 0.25,\, 0.5] & 0.144\\

\bottomrule
\end{tabular*}
\end{table}




\section{Results}

\subsection{\textit{Effect of Mass Ratio}}

The power coefficients reported in this section extend beyond the conventional hydrodynamic pitch-power metric by separating the actuator-side pitch-power balance into three contributions. In the present semi-passive configuration, only the pitching motion is externally actuated, whereas the heaving motion responds passively to fluid loading and structural inertia. The first contribution is the \textit{hydrodynamic pitch power}, ${C}_{P,\theta}^{\,\mathrm{hydro}}(t)$, associated with the fluid torque $\tau_{\theta}^{\,\mathrm{fluid}}(t)$ acting on the prescribed pitching motion. The second is the added rotational inertial contribution from $I_\mathrm{virtual}\ddot{\theta}(t)$, which does not contribute to the cycle-averaged hydrodynamic efficiency but can alter the instantaneous actuator demand. The third is the added heave-induced contribution from $S^\mathrm{virt}_\mathrm{foil}\ddot{h}(t)\cos\theta(t)$, which couples passive heave acceleration to the prescribed pitching motion~\cite{sorensen2024developing,veilleux2017numerical}. Unlike the hydrodynamic contribution, the latter two terms scale directly with the prescribed virtual foil mass. They therefore provide the mechanism by which changing $m_\mathrm{foil}^*$ can alter the actuator-level pitch-power demand, even when the prescribed pitching kinematics and incoming flow are held fixed.

\begin{figure}
	\centering
        \begin{subfigure}{1.0\textwidth}
            \centering
	       \includegraphics[width=\linewidth]{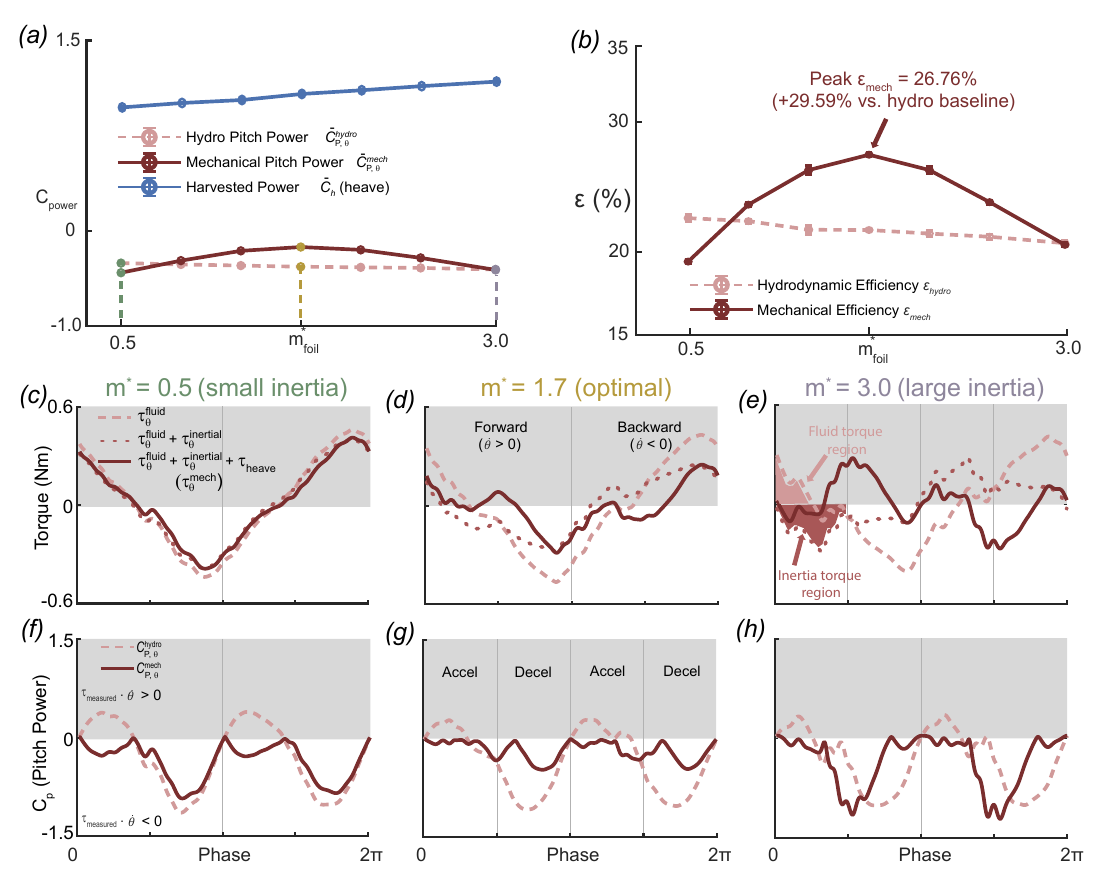}
        \end{subfigure}
    
	\caption{Cycle-averaged pitch-power coefficients $\bar{C}_{P,\theta}^{\,\mathrm{hydro}}$ and  $\bar{C}_{P,\theta}^{\,\mathrm{mech}}$ together with harvested heave power $\bar{C}_h$ (a) and efficiencies $\varepsilon_{\mathrm{hydro}}$, $\varepsilon_{\mathrm{mech}}$ (b) versus foil mass ratio for a leading-edge ($x/c = 0$) hydrofoil at $\kappa = 0.144$. Instantaneous fluid, fluid and inertial, and mechanical torques $\tau_{\theta}^{\,\mathrm{fluid}}(t)$, $\tau_{\theta}^{\,\mathrm{fluid}}(t)+\tau_{\theta}^{\,\mathrm{inertial}}(t)$, $\tau_{\theta}^{\,\mathrm{mech}}(t)$ (c, d, e) and instantaneous hydrodynamic, mechanical pitch power coefficients $C_{P,\theta}^{\,\mathrm{hydro}}(t)$,  $C_{P,\theta}^{\,\mathrm{mech}}(t)$ (f, g, h) for representative foil mass ratios ($m_\mathrm{foil}^*$). \protect}
	\label{FIG:2}
\end{figure}
Figure~\ref{FIG:2}(a) summarizes these power coefficients across the tested mass ratios, with the prescribed pitching amplitude, frequency, waveform, and incoming flow held fixed. The hydrodynamic pitch power remains nearly invariant at approximately $-0.349$ across all $m_\mathrm{foil}^*$, indicating that the fluid torque about the pivot is only weakly affected by inertia-induced changes in the passive heave response. The harvested heave power, by contrast, increases gradually with $m_\mathrm{foil}^*$, consistent with the broader observation of Deng et al.~\cite{deng2015inertial} that mass ratio alters the response and power balance of semi-passive flapping foils. The mechanical pitch-power consumption exhibits a much stronger and qualitatively different response: it decreases first, reaches a minimum cost of $-0.188$ at $m_\mathrm{foil}^* = 1.7$---about $46.13\%$ smaller in magnitude than the hydrodynamic pitch-power baseline---and then increases again. This non-monotonic trend reflects two competing effects of inertia: increasing mass can redistribute kinetic energy within a cycle, but it also increases the actuation effort required during acceleration.

Figure~\ref{FIG:2}(b) translates this power balance into efficiencies. The hydrodynamic efficiency remains nearly flat with $m_\mathrm{foil}^*$ in the present experiments, mirroring the weak dependence of the hydrodynamic pitch power on mass ratio. The mechanical efficiency, however, rises to a maximum of $26.76\%$ at $m_\mathrm{foil}^* = 1.7$ before declining, corresponding to a $29.59\%$ improvement over the hydrodynamic baseline for the same configuration: a leading-edge pivot and reduced frequency $\kappa = 0.144$. This peak results from the synchronized behavior of the two terms entering the efficiency numerator: the mechanical pitch-power demand reaches its minimum at intermediate $m_\mathrm{foil}^*$, while the harvested heave power continues to increase gradually across the tested range. 
Because the available flow power, $0.5\rho U_\infty^3 A_s$, varies only weakly with $m_\mathrm{foil}^*$ through small changes in swept area, the efficiency maximum is controlled primarily by the balance between harvested heave power, $\langle L_\mathrm{fluid}(t)\dot{h}(t)\rangle$, and mechanical pitch-actuation cost, $\langle|\tau_{\theta}^{\,\mathrm{mech}}(t)\dot{\theta}(t)|\rangle$. Beyond the optimum, the increasing mechanical pitch-power demand outpaces the continued rise in harvested heave power, causing $\varepsilon_{\mathrm{mech}}$ to decrease.

To understand the non-monotonic behavior of the mechanical pitch power, we first isolate the role of the pitch inertial term $\tau_{\theta}^{\,\mathrm{inertial}}(t) = I_\mathrm{virtual}\ddot{\theta}(t)$ in the instantaneous profiles of three representative cases: $m_\mathrm{foil}^* = 0.5$, $1.7$, and $3.0$ (Fig.~\ref{FIG:2}(c--e)). Increasing $m_\mathrm{foil}^*$ does not simply raise the actuator demand uniformly throughout the cycle; instead, the growing inertial torque redistributes the demand in a phase-dependent manner. For example, in the large-inertia case $m_\mathrm{foil}^* = 3.0$ (Fig.~\ref{FIG:2}(e)), $I_\mathrm{virtual}\ddot{\theta}(t)$ enlarges the torque-demand region during acceleration phases relative to the hydrodynamic baseline, as indicated by the dark and light shaded areas. During deceleration phases, the kinetic energy stored earlier in the cycle is returned to the pitching motion. This inertial release does not remove the fluid torque, but it can reduce the net torque that must be supplied or dissipated by the actuator depending on its phase relative to the hydrodynamic loading. In this sense, inertia acts as a within-cycle energy reservoir, redistributing actuator demand between phases without substantially altering the net fluid input~\cite{siala2020experimental}.

We next include the heave-induced coupling term, $\tau_{\mathrm{heave}}(t) = S^\mathrm{virt}_\mathrm{foil}\ddot{h}(t)\cos\theta(t)$, which introduces a second phase-dependent contribution to the actuator demand. Unlike the pitch inertial term, whose phase is fixed by the prescribed pitching kinematics, the phase of $S^\mathrm{virt}_\mathrm{foil}\ddot{h}(t)\cos\theta(t)$ is governed by the passive heave acceleration $\ddot{h}(t)$. Its phase relative to the pitching motion is therefore governed by the passive heave response, which is sensitive to heave--pitch coupling and pivot geometry~\cite{boudreau2018experimental,mackowski2017effect,veilleux2017numerical}.  Under the resonance condition imposed here, $f/f_n = 1$, the passive heave displacement $h(t)$ lags the pitch angle $\theta(t)$ by approximately $\pi/2$. Consequently, $\ddot{h}(t)$ leads $\theta(t)$ by approximately $\pi/2$, whereas $I_\mathrm{virtual}\ddot{\theta}(t)$ lags $\theta(t)$ by $\pi$. The heave-induced torque is therefore offset by approximately $\pi/2$ from the pitch inertial torque, allowing the two contributions to partially cancel over substantial portions of the cycle. At the optimal mass ratio (Fig.~\ref{FIG:2}(d)), this cancellation suppresses the minimum pitch torque-demand region over each cycle. For the over-heavy foil (Fig.~\ref{FIG:2}(e)), however, both the rotational inertial term and the heave-induced coupling term become large. The coupling term no longer acts primarily as a compensating contribution; instead, its own torque excursions add substantially to the actuator-side load over portions of the cycle. The net torque-demand region therefore increases again, imposing a larger mechanical pitch-power cost.

Together, these phase-dependent inertial and heave-induced contributions reduce the mechanical pitch-power cost in the present leading-edge configuration, producing the efficiency peak at $m_\mathrm{foil}^* = 1.7$ (Fig.~\ref{FIG:2}(b)). Because the phase and magnitude of the heave-induced torque depend on geometry, this favorable suppression mechanism is not guaranteed in general. The following subsection tests this point by varying the pitching-axis location.

\subsection{\textit{Effect of pitching-axis location}}

The pitching-axis location is qualitatively different from the foil mass ratio $m^{*}_{\mathrm{foil}}$ examined in Section~3.1. Whereas $m^{*}_{\mathrm{foil}}$ scales the magnitude of inertial loading, the pivot location controls three coupled geometric quantities at once: the lever arm of the fluid torque, the static moment $S_{\mathrm{foil}} = m_{\mathrm{foil}}r$ governing heave--pitch coupling~\cite{boudreau2018experimental,veilleux2017numerical}, and the rotational inertia $I_\mathrm{physical}$ and $I_\mathrm{virtual}$ about the pivot~\cite{qadri2020fluid}. As the pivot moves from the leading edge toward the center of mass (COM), the static moment decreases and eventually vanishes, while the fluid moment arm also changes substantially. These geometric changes alter not only the magnitude but also the phase---and in some cases the sign---of the dominant torque contributions. The pivot location therefore acts as a directional controller of the inertia effect, complementing the role of $m^{*}_{\mathrm{foil}}$ as a magnitude controller. 

\begin{figure}
	\centering
        \begin{subfigure}{1.0\textwidth}
            \centering
	       \includegraphics[width=\linewidth]{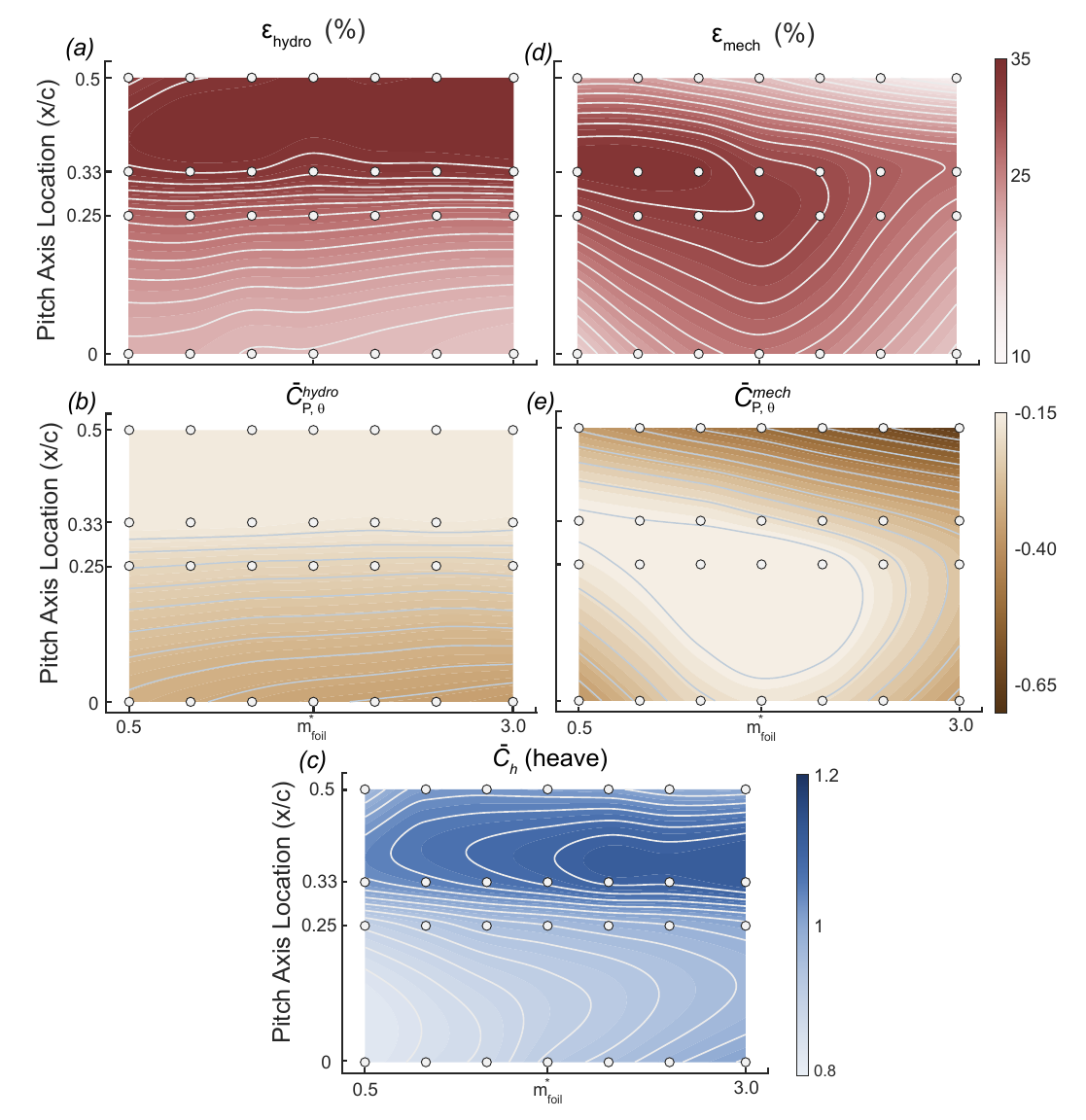}
        \end{subfigure}
    
	\caption{Contour maps of (a) hydrodynamic efficiency $\varepsilon_{\mathrm{hydro}}$, (b) hydrodynamic pitch-power coefficient $\bar{C}_{P,\theta}^{\,\mathrm{hydro}}$, (c) heave power coefficient $\bar{C}_{h}$, (d) mechanical efficiency $\varepsilon_{\mathrm{mech}}$, and (e) mechanical pitch-power coefficient $\bar{C}_{P,\theta}^{\,\mathrm{mech}}$, with respect to foil mass ratio $m^*_{\mathrm{foil}}$ and pitching-axis location $x/c$ at reduced frequency $\kappa = 0.144$. White markers indicate the experimental test conditions used to construct the contour distributions.}
	\label{FIG:3}
\end{figure}

Figure~\ref{FIG:3} summarizes how foil mass ratio and pivot location jointly shape efficiency and pitch power at $\kappa = 0.144$. The hydrodynamic and mechanical efficiency maps identify different optimal regions. The hydrodynamic efficiency $\varepsilon_{\mathrm{hydro}}$ (Fig.~\ref{FIG:3}(a)) is governed primarily by pivot location, increasing from approximately $21\%$ at the leading edge to approximately $34\%$ at mid-chord, with only weak dependence on $m^{*}_{\mathrm{foil}}$. This trend results from two pivot-driven effects: as the pivot shifts aft, the hydrodynamic pitch-power magnitude $|\bar{C}_{P,\theta}^{\,\mathrm{hydro}}|$ decreases from approximately $0.35$ to $0.17$ because of the reduced fluid-torque moment arm (Fig.~\ref{FIG:3}(b)), while the harvested heave power $\bar{C}_h$ increases (Fig.~\ref{FIG:3}(c)). Thus, the hydrodynamic efficiency map is structured mainly by pivot geometry rather than foil inertia.

The mechanical efficiency $\varepsilon_{\mathrm{mech}}$ (Fig.~\ref{FIG:3}(d)) shows a different optimum. It reaches a global maximum of $33.96\%$ near $(x/c,\,m^{*}_{\mathrm{foil}}) = (0.33,\,0.9)$ and a minimum of $13.70\%$ at $(0.5,\,3.0)$. Compared with $\varepsilon_{\mathrm{hydro}}$, $\varepsilon_{\mathrm{mech}}$ is higher for upstream-of-COM pivots, approximately coincident near $x/c \approx 0.33$, while up to $59.71\%$ lower at mid-chord for $m^{*}_{\mathrm{foil}} = 3.0$. The largest \textit{relative} improvement remains at the leading edge ($+29.59\%$ at $m^{*}_{\mathrm{foil}} = 1.7$), whereas the highest \textit{absolute} mechanical efficiency occurs near one-third chord, where the hydrodynamic baseline is already high and the inertial penalty is small. Consistently, the mechanical pitch-power magnitude in Fig.~\ref{FIG:3}(e) reaches its minimum, $|\bar{C}_{P,\theta}^{\,\mathrm{mech}}| = 0.106$, at $(x/c,\,m^{*}_{\mathrm{foil}}) = (0.25,\,1.7)$ and increases to $0.67$ at $(0.5,\,3.0)$.

\begin{figure}
	\centering
        \begin{subfigure}{1.0\textwidth}
            \centering
	       \includegraphics[width=\linewidth]{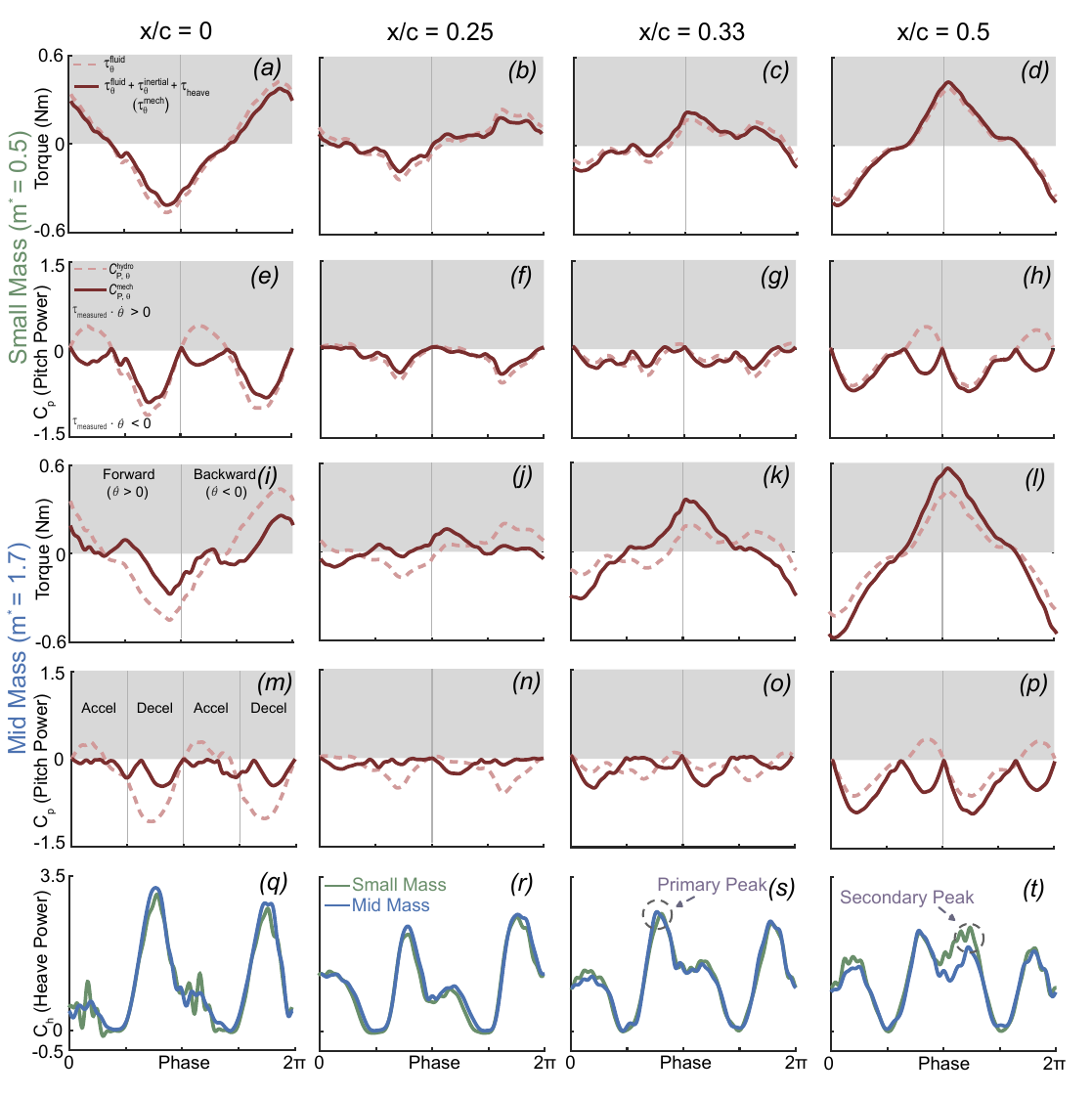}
        \end{subfigure}
    
	\caption{Instantaneous fluid torque $\tau_{\theta}^{\,\mathrm{fluid}}(t)$, mechanical torque $\tau_{\theta}^{\,\mathrm{mech}}(t)$, hydrodynamic pitch power $C_{P,\theta}^{\,\mathrm{hydro}}(t)$, and mechanical pitch power $C_{P,\theta}^{\,\mathrm{mech}}(t)$ as functions of phase for pitching-axis locations $x/c \in \{0,\,0.25,\,0.33,\,0.5\}$ at small-mass, $m^{*}_{\mathrm{foil}} = 0.5$ (a--h) and mid-mass, $m^{*}_{\mathrm{foil}} = 1.7$ (i--p) configurations with $\kappa = 0.144$. Panels (q--t) show the corresponding instantaneous heave power $C_h(t)$. The heave power remains relatively insensitive to mass-ratio variations at fixed $\kappa$.}
	\label{FIG:4}
\end{figure}

Figure~\ref{FIG:4} resolves the contour trends into instantaneous, phase-dependent contributions across two representative mass ratios. The small-mass row (Fig.~\ref{FIG:4}(a--h)) provides a baseline in which inertia-induced redistribution is weak and the fluid and mechanical torque curves remain close. Even in this limit, pivot location substantially reshapes the fluid-torque waveform: as the pivot shifts from the leading edge to mid-chord, the torque first decreases in magnitude at $x/c=0.25$, then advances in phase near $x/c=0.33$, and finally becomes nearly inverted at $x/c=0.5$, with a phase advance of approximately $160^\circ$. These waveform changes directly shape the actuator pitch-power consumption (Fig.~\ref{FIG:4}(e--h)), which is smallest for $x/c=0.25$--$0.33$ and largest at $x/c=0$ and $0.5$.

The corresponding heave-power responses (Fig.~\ref{FIG:4}(q--t)) show a systematic waveform transition as the pivot moves from the leading edge to mid-chord. At the leading edge, each half-cycle is dominated by a single narrow primary peak. Moving the pivot toward the quarter- and one-third-chord positions weakens this primary peak while amplifying a secondary peak on its descending side. By mid-chord, the secondary peak becomes comparable to the primary peak, and the two merge into a broad plateau. Thus, pivot location reshapes not only the pitch-torque demand but also the timing and distribution of harvested heave power within each cycle.

Throughout this evolution, the dominant peak phases remain anchored near $\phi \approx 3\pi/4$ and $7\pi/4$, indicating that resonance fixes the principal phase relationship between fluid lift and heave velocity largely independently of pivot geometry. Comparable behavior has been observed in both \textit{fully passive}~\cite{qadri2019experimental} and \textit{fully active}~\cite{su2019confinement} energy harvesters, suggesting that the timing of the principal energy-transfer events is governed primarily by the resonant dynamics of the system rather than by the specific actuation configuration. The progressive rise of the secondary peak indicates an increasingly important second unsteady-fluid-forcing event within each half-cycle as the pivot approaches the COM. This flow-structure mechanism is examined further using the PIV measurements later in this Section.

At the intermediate mass ratio, $m^{*}_{\mathrm{foil}} = 1.7$ (Fig.~\ref{FIG:4}(i--p)), inertia-induced redistribution becomes strong enough to modify the pitch energy balance. For upstream-of-COM pivots, the favorable mechanism identified in Section~3.1 persists: added inertia reduces the mechanical torque-demand region, and the smaller hydrodynamic pitch torque at $x/c=0.25$--$0.33$ further lowers the cycle-averaged mechanical pitch-power cost. This produces the global minimum in Fig.~\ref{FIG:3}(e), $|\bar{C}_{P,\theta}^{\,\mathrm{mech}}| = 0.106$, at $(x/c,\,m^{*}_{\mathrm{foil}}) = (0.25,\,1.7)$.

The mid-chord case (Fig.~\ref{FIG:4}(l,p)) departs from this favorable pattern. Because the fluid torque is nearly inverted relative to the leading-edge waveform, the inertial contribution that assists upstream-of-COM pivots instead opposes the prescribed motion over substantial portions of the cycle. The mechanical pitch-power magnitude at $m^{*}_{\mathrm{foil}} = 1.7$ therefore exceeds the small-mass baseline, and this penalty grows at higher mass ratios, leading to the $\varepsilon_{\mathrm{mech}}$ minimum of $13.70\%$ at $(x/c,\,m^{*}_{\mathrm{foil}}) = (0.5,\,3.0)$ in Fig.~\ref{FIG:3}(d).

The near overlap of the small- and intermediate-mass heave-power traces in Fig.~\ref{FIG:4}(q--t) shows that harvested heave power is largely insensitive to $m^{*}_{\mathrm{foil}}$ at fixed pivot location and reduced frequency. The cross-pivot waveform evolution is therefore governed primarily by pivot geometry, whereas the strong $m^{*}_{\mathrm{foil}}$ dependence of $\varepsilon_{\mathrm{mech}}$ arises mainly from the pitch side of the energy balance.

\begin{figure}
	 \centering
        \begin{subfigure}{1.0\textwidth}
            \centering
	       \includegraphics[width=\linewidth]{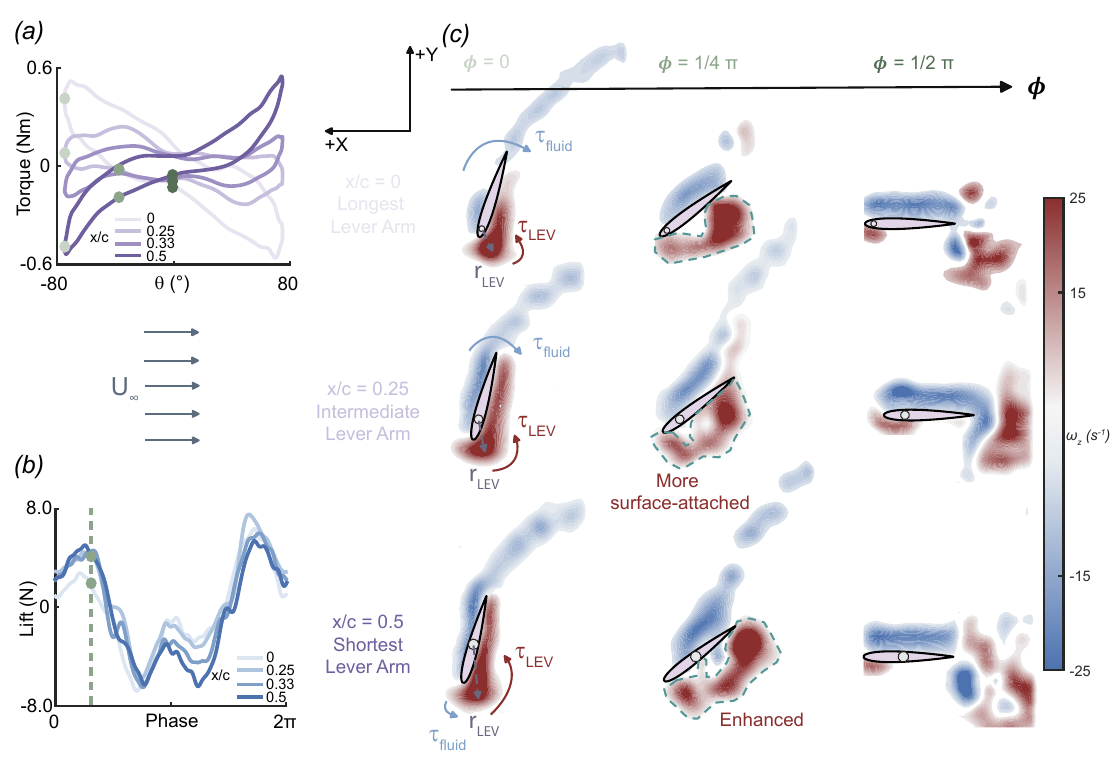}
        \end{subfigure}
	\caption{Effect of pitch-axis location on the flow field and force/torque response at $m^{*}_{\mathrm{foil}}=1.7$ and $\kappa=0.144$: (a) fluid torque $\tau_{\theta}^{\,\mathrm{fluid}}$ versus pitch angle, showing the pitch-axis-dependent torque loops; (b) instantaneous lift force $L_\mathrm{{fluid}}(t)$ over one motion cycle; (c) phase-resolved vortex topology at three representative phases ($\phi = 0$, $\pi/4$, $\pi/2$) for the leading-edge, quarter-chord, and mid-chord configurations, highlighting the leading-edge vortex (LEV) and its feeding shear layer. \protect}
	\label{FIG:5}
\end{figure}

To trace the pitch-axis trends in hydrodynamic pitch power back to their flow-field origin, we examine the fluid torque $\tau_{\theta}^{\,\mathrm{fluid}}(t)$ through the torque--pitch-angle relationship and the phase-resolved vorticity fields (Fig.~\ref{FIG:5}(a,c)). The torque loops provide a phase-resolved view of the fluid--structure interaction, where loop shape, enclosed area, and phase asymmetry reflect the evolution of unsteady loading over the oscillation cycle.

Figure~\ref{FIG:5}(a) shows that both the magnitude and direction of the measured torque change substantially as the pitching axis shifts from the leading edge to mid-chord. This reversal can be interpreted from the competition between distributed surface-pressure loading and the concentrated suction associated with the leading-edge vortex (LEV). At the leading edge (Fig.~\ref{FIG:5}(c), $\phi=0$), the surface-pressure loading acts through a long moment arm and produces a large clockwise torque, whereas the LEV forms close to the pivot and contributes only weakly in the opposite direction. At mid-chord, the fore--aft pressure contributions largely cancel about the pivot, while the LEV lies upstream of the pivot and acts through a longer moment arm, producing a counter-clockwise torque. This shift in the dominant moment arm explains the inverted torque loop observed for the mid-chord configuration in Fig.~\ref{FIG:5}(a).

The pitch-axis location also reshapes the lift response (Fig.~\ref{FIG:5}(b)). Although the LEV core that forms during the early acceleration phase is of comparable strength across all configurations, the feeding shear layer connecting the leading edge to this core becomes progressively more extensive as the pivot moves aft toward mid-chord (Fig.~\ref{FIG:5}(c)). Because this shear layer represents bound vorticity attached to the suction surface, its growth increases the near-surface circulation around the foil; by the Kutta--Joukowski relation, $L = \rho U_\infty \Gamma$, the enhanced and more sustained circulation maintains elevated lift later in the cycle. This is consistent with the secondary lift peak near $\phi \approx \pi/4$ in Fig.~\ref{FIG:5}(b), which---combined with the heave velocity---produces the secondary peak in the heave-power waveform shown in Fig.~\ref{FIG:4}(q--t). The pitch-axis location therefore exerts a dual influence: it reduces and ultimately reverses the effective torque lever arm, while simultaneously sustaining circulation that enhances lift during the later stages of the cycle.

\subsection{\textit{Effect of reduced frequency}}

The reduced frequency, $\kappa = fc/U_{\infty}$, plays a central role in setting the overall energy balance of the system. Whereas the foil mass ratio $m^{*}_{\mathrm{foil}}$ scales the magnitude of inertial loading (Section~3.1) and the pitching-axis location redirects it geometrically (Section~3.2), $\kappa$ amplifies the inertial torque dynamically through the $(2\pi f)^2$ dependence of angular acceleration. At the same time, increasing $\kappa$ strengthens the unsteady fluid forcing. The competition between these two frequency-dependent effects determines the optimal operating condition.

\begin{figure} 
    \centering
        \begin{subfigure}{1.0\textwidth}
            \centering
	       \includegraphics[width=\linewidth]{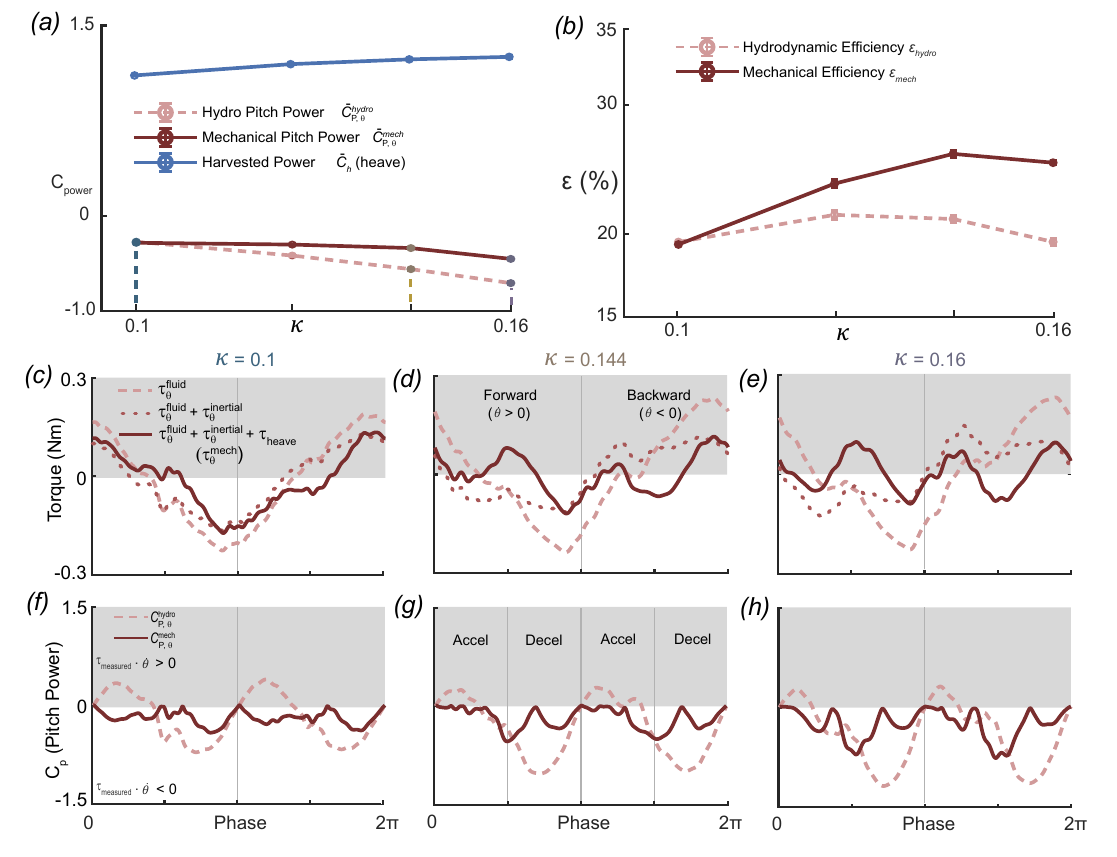}
        \end{subfigure}
    
	\caption{Cycle-averaged pitch-power coefficients $\bar{C}_{P,\theta}^{\,\mathrm{hydro}}$ and $\bar{C}_{P,\theta}^{\,\mathrm{mech}}$ together with harvested heave power $\bar{C}_h$ (a) and efficiencies $\varepsilon_{\mathrm{hydro}}$ and $\varepsilon_{\mathrm{mech}}$ (b) versus reduced frequency for a leading-edge hydrofoil ($x/c = 0$) at $m^{*}_{\mathrm{foil}} = 2.1$. Instantaneous fluid, fluid and inertial, and mechanical torques $\tau_{\theta}^{\,\mathrm{fluid}}(t)$, $\tau_{\theta}^{\,\mathrm{fluid}}(t) + \tau_{\theta}^{\,\mathrm{inertial}}(t)$, and $\tau_{\theta}^{\,\mathrm{mech}}(t)$ are shown in (c--e), and instantaneous hydrodynamic and mechanical pitch powers $C_{P,\theta}^{\,\mathrm{hydro}}(t)$ and $C_{P,\theta}^{\,\mathrm{mech}}(t)$ are shown in (f--h), for representative reduced frequencies $\kappa$.}
	\label{FIG:6}
\end{figure}

Figure~\ref{FIG:6}(a) summarizes the cycle-averaged power coefficients across the tested reduced-frequency range for the leading-edge pivot at $m^{*}_{\mathrm{foil}} = 2.1$. All three coefficients increase in magnitude with $\kappa$, but at different rates. The harvested heave power $\bar{C}_h$ rises modestly, whereas the hydrodynamic pitch-power magnitude grows more steeply as the faster pitching motion strengthens the unsteady fluid forcing. Similar reduced-frequency sensitivity has been reported in previous semi-passive and prescribed-motion oscillating-foil studies~\cite{deng2015inertial,zhu2011optimal}. In contrast, the mechanical pitch-power magnitude $|\bar{C}_{P,\theta}^{\,\mathrm{mech}}|$ grows more slowly than the hydrodynamic pitch-power baseline: at $\kappa = 0.1$, the two pitch-power magnitudes are closely matched at approximately $0.17$, whereas by $\kappa = 0.16$, the hydrodynamic magnitude has increased by roughly a factor of five and the mechanical magnitude by less than a factor of three.

These different growth rates produce distinct efficiency trends (Fig.~\ref{FIG:6}(b)). At $\kappa = 0.10$, the hydrodynamic and mechanical efficiencies are nearly indistinguishable, at $18.71\%$ and $18.51\%$, respectively, because the inertial torque is too small to separate the two metrics. As $\kappa$ increases, the curves diverge: $\varepsilon_{\mathrm{hydro}}$ peaks mildly at $\kappa = 0.125$ and then decreases slightly, whereas $\varepsilon_{\mathrm{mech}}$ reaches its maximum of $25.48\%$ at $\kappa = 0.144$ before declining modestly at $\kappa = 0.16$. Thus, the optimal reduced-frequency range for mechanical efficiency is shifted toward $\kappa = 0.144$, where the gain in harvested heave power and the reduction in effective mechanical pitch cost are best balanced.  This optimum falls within the reduced-frequency range, $\kappa \approx 0.125$--$0.16$, commonly reported for energy-harvesting flapping foils~\cite{kinsey2008parametric,siala2020experimental,zhu2011optimal}.

The instantaneous torque decomposition (Fig.~\ref{FIG:6}(c--e)) explains why $|\bar{C}_{P,\theta}^{\,\mathrm{mech}}|$ grows more slowly than $|\bar{C}_{P,\theta}^{\,\mathrm{hydro}}|$. The fluid torque $\tau_\theta^{\,\mathrm{fluid}}(t)$ retains a similar waveform across the three reduced frequencies, with moderate amplitude growth as the unsteady loading strengthens. By contrast, the inertial torque $I_\mathrm{virtual}\ddot{\theta}(t)$ scales with $(2\pi f)^2$, amplifying the phase-dependent energy redistribution described in Section~3.1. In the present leading-edge configuration, this inertial contribution combines with the heave-induced coupling torque, $\tau_{\mathrm{heave}}(t)=S^\mathrm{virt}_\mathrm{foil}\ddot{h}(t)\cos\theta(t)$, to reduce the mechanical torque excursions over substantial portions of the cycle. This effect is weak at $\kappa=0.1$ (Fig.~\ref{FIG:6}(c)), strongest near $\kappa=0.144$--$0.16$ (Fig.~\ref{FIG:6}(d,e)), where the mechanical torque remains visibly closer to zero than the fluid torque. Thus, increasing $\kappa$ at fixed mass amplifies the same inertia-induced redistribution mechanism identified in Section~3.1.

The instantaneous pitch-power profiles (Fig.~\ref{FIG:6}(f--h)) translate this torque redistribution into actuator cost. As $\kappa$ increases, the hydrodynamic pitch power develops larger negative excursions, reflecting the growing fluid torque required to sustain the prescribed pitching motion. The mechanical pitch power also increases in magnitude, but its excursions remain smaller because the inertial and heave-induced contributions partially reduce the net actuator demand over favorable phases. This explains why $|\bar{C}_{P,\theta}^{\,\mathrm{mech}}|$ grows more slowly than $|\bar{C}_{P,\theta}^{\,\mathrm{hydro}}|$ in Fig.~\ref{FIG:6}(a). Beyond $\kappa=0.144$, however, the inertia-amplified torque excursions begin to outpace the gain in harvested heave power, leading to the modest decrease in $\varepsilon_{\mathrm{mech}}$ at $\kappa=0.16$.

\section{Conclusions}

The present study demonstrates that the harvesting performance of a semi-passive oscillating hydrofoil is governed by the coupled interaction among unsteady fluid forcing, rotational inertia, and heave--pitch coupling. Unlike conventional hydrodynamic analyses that focus on fluid--foil power transfer, the actuator-level framework used here shows that the net mechanical energy balance depends critically on the phase relationship among fluid torque, inertial loading, and passive heave response. Foil inertia acts as a within-cycle energy reservoir, storing kinetic energy during acceleration and releasing it during deceleration; depending on its synchronization with the fluid forcing and heave-induced coupling, this redistribution can either reduce or increase the mechanical pitch-power cost. Pitching-axis location further controls this balance by changing the effective hydrodynamic moment arm, which modifies both the magnitude and direction of the fluid-induced torque. Phase-resolved PIV measurements show that similar leading-edge vortices can produce substantially different torque responses because the vortex-induced force acts through different moment arms relative to the pitching axis. Reduced frequency provides a third control parameter by simultaneously strengthening unsteady fluid forcing and amplifying inertia-related actuator demand. Within the investigated parameter space, the most favorable performance occurred for $\kappa \approx 0.125$--$0.16$, quarter-chord to one-third-chord pitching axes, and relatively low foil mass ratios. The highest mechanical efficiency, $33.96\%$, was achieved near the one-third-chord pivot, whereas the mechanical efficiency dropped to $13.70\%$ at the mid-chord configuration despite its high hydrodynamic efficiency.

From an engineering perspective, these results show that optimal harvesting performance is achieved through coordinated tuning of structural and operating parameters rather than by maximizing hydrodynamic efficiency alone. Mechanical efficiency should therefore serve as a primary design metric for non-regenerative oscillating-foil energy harvesters, where energy associated with prescribed pitching motion cannot be recovered through a generator. The present results identify a practical design window in which vortex-induced energy extraction remains strong while actuator-level power consumption is minimized. Although the experiments were conducted under steady inflow conditions, practical hydrokinetic environments involve time-varying velocity and turbulence; future work could focus on adaptive control strategies that maintain favorable phase synchronization under changing flow conditions. Regenerative pitch-drive architectures also offer a promising path for recovering part of the braking energy otherwise dissipated during deceleration, thereby further improving the net energy-conversion efficiency of oscillating-foil harvesters.

\printcredits

\begin{thebibliography}{44}
\expandafter\ifx\csname natexlab\endcsname\relax\def\natexlab#1{#1}\fi
\providecommand{\url}[1]{\texttt{#1}}
\providecommand{\href}[2]{#2}
\providecommand{\path}[1]{#1}
\providecommand{\DOIprefix}{doi:}
\providecommand{\ArXivprefix}{arXiv:}
\providecommand{\URLprefix}{URL: }
\providecommand{\Pubmedprefix}{pmid:}
\providecommand{\doi}[1]{\href{http://dx.doi.org/#1}{\path{#1}}}
\providecommand{\Pubmed}[1]{\href{pmid:#1}{\path{#1}}}
\providecommand{\bibinfo}[2]{#2}
\ifx\xfnm\relax \def\xfnm[#1]{\unskip,\space#1}\fi
\bibitem[{Ashraf et~al.(2011)Ashraf, Young, Lai and Platzer}]{ashraf2011numerical}
\bibinfo{author}{Ashraf, M.}, \bibinfo{author}{Young, J.}, \bibinfo{author}{Lai, J.}, \bibinfo{author}{Platzer, M.}, \bibinfo{year}{2011}.
\newblock \bibinfo{title}{Numerical analysis of an oscillating-wing wind and hydropower generator}.
\newblock \bibinfo{journal}{AIAA journal} \bibinfo{volume}{49}, \bibinfo{pages}{1374--1386}.
\bibitem[{Boudis et~al.(2021)Boudis, Oualli, Benzaoui, Guerri, Bayeul-Lain{\'e} and Coutier-Delgosha}]{boudis2021effects}
\bibinfo{author}{Boudis, A.}, \bibinfo{author}{Oualli, H.}, \bibinfo{author}{Benzaoui, A.}, \bibinfo{author}{Guerri, O.}, \bibinfo{author}{Bayeul-Lain{\'e}, C.}, \bibinfo{author}{Coutier-Delgosha, O.}, \bibinfo{year}{2021}.
\newblock \bibinfo{title}{Effects of non-sinusoidal motion and effective angle of attack on energy extraction performance of a fully-activated flapping foil}.
\newblock \bibinfo{journal}{Journal of Applied Fluid Mechanics} \bibinfo{volume}{14}, \bibinfo{pages}{485--498}.
\bibitem[{Boudreau et~al.(2018)Boudreau, Dumas, Rahimpour and Oshkai}]{boudreau2018experimental}
\bibinfo{author}{Boudreau, M.}, \bibinfo{author}{Dumas, G.}, \bibinfo{author}{Rahimpour, M.}, \bibinfo{author}{Oshkai, P.}, \bibinfo{year}{2018}.
\newblock \bibinfo{title}{Experimental investigation of the energy extraction by a fully-passive flapping-foil hydrokinetic turbine prototype}.
\newblock \bibinfo{journal}{Journal of Fluids and Structures} \bibinfo{volume}{82}, \bibinfo{pages}{446--472}.
\bibitem[{Boudreau et~al.(2020)Boudreau, Picard-Deland and Dumas}]{boudreau2020parametric}
\bibinfo{author}{Boudreau, M.}, \bibinfo{author}{Picard-Deland, M.}, \bibinfo{author}{Dumas, G.}, \bibinfo{year}{2020}.
\newblock \bibinfo{title}{A parametric study and optimization of the fully-passive flapping-foil turbine at high reynolds number}.
\newblock \bibinfo{journal}{Renewable Energy} \bibinfo{volume}{146}, \bibinfo{pages}{1958--1975}.
\bibitem[{Deng et~al.(2014)Deng, Caulfield and Shao}]{deng2014effect}
\bibinfo{author}{Deng, J.}, \bibinfo{author}{Caulfield, C.}, \bibinfo{author}{Shao, X.}, \bibinfo{year}{2014}.
\newblock \bibinfo{title}{Effect of aspect ratio on the energy extraction efficiency of three-dimensional flapping foils}.
\newblock \bibinfo{journal}{Physics of Fluids} \bibinfo{volume}{26}.
\bibitem[{Deng et~al.(2015)Deng, Teng, Pan and Shao}]{deng2015inertial}
\bibinfo{author}{Deng, J.}, \bibinfo{author}{Teng, L.}, \bibinfo{author}{Pan, D.}, \bibinfo{author}{Shao, X.}, \bibinfo{year}{2015}.
\newblock \bibinfo{title}{Inertial effects of the semi-passive flapping foil on its energy extraction efficiency}.
\newblock \bibinfo{journal}{Physics of Fluids} \bibinfo{volume}{27}.
\bibitem[{Duarte et~al.(2019)Duarte, Dellinger, Dellinger, Ghenaim and Terfous}]{duarte2019experimental}
\bibinfo{author}{Duarte, L.}, \bibinfo{author}{Dellinger, N.}, \bibinfo{author}{Dellinger, G.}, \bibinfo{author}{Ghenaim, A.}, \bibinfo{author}{Terfous, A.}, \bibinfo{year}{2019}.
\newblock \bibinfo{title}{Experimental investigation of the dynamic behaviour of a fully passive flapping foil hydrokinetic turbine}.
\newblock \bibinfo{journal}{Journal of Fluids and Structures} \bibinfo{volume}{88}, \bibinfo{pages}{1--12}.
\bibitem[{Duarte et~al.(2021)Duarte, Dellinger, Dellinger, Ghenaim and Terfous}]{duarte2021experimental}
\bibinfo{author}{Duarte, L.}, \bibinfo{author}{Dellinger, N.}, \bibinfo{author}{Dellinger, G.}, \bibinfo{author}{Ghenaim, A.}, \bibinfo{author}{Terfous, A.}, \bibinfo{year}{2021}.
\newblock \bibinfo{title}{Experimental optimisation of the pitching structural parameters of a fully passive flapping foil turbine}.
\newblock \bibinfo{journal}{Renewable Energy} \bibinfo{volume}{171}, \bibinfo{pages}{1436--1444}.
\bibitem[{Handy-Cardenas et~al.(2025)Handy-Cardenas, Zhu and Breuer}]{handy2025optimal}
\bibinfo{author}{Handy-Cardenas, E.E.}, \bibinfo{author}{Zhu, Y.}, \bibinfo{author}{Breuer, K.S.}, \bibinfo{year}{2025}.
\newblock \bibinfo{title}{Optimal kinematics for energy harvesting using favourable wake--foil interactions in tandem oscillating hydrofoils}.
\newblock \bibinfo{journal}{Journal of Fluid Mechanics} \bibinfo{volume}{1012}, \bibinfo{pages}{A23}.
\bibitem[{Jones et~al.(2003)Jones, Lindsey and Platzer}]{jones2003investigation}
\bibinfo{author}{Jones, K.D.}, \bibinfo{author}{Lindsey, K.}, \bibinfo{author}{Platzer, M.}, \bibinfo{year}{2003}.
\newblock \bibinfo{title}{An investigation of the fluid-structure interaction in an oscillating-wing micro-hydropower generator} .
\bibitem[{Karakas and Fenercioglu(2016)}]{karakas2016effect}
\bibinfo{author}{Karakas, F.}, \bibinfo{author}{Fenercioglu, I.}, \bibinfo{year}{2016}.
\newblock \bibinfo{title}{Effect of side-walls on flapping-wing power-generation: An experimental study}.
\newblock \bibinfo{journal}{Journal of Applied Fluid Mechanics} \bibinfo{volume}{9}, \bibinfo{pages}{2769--2779}.
\bibitem[{Kim et~al.(2017)Kim, Strom, Mandre and Breuer}]{kim2017energy}
\bibinfo{author}{Kim, D.}, \bibinfo{author}{Strom, B.}, \bibinfo{author}{Mandre, S.}, \bibinfo{author}{Breuer, K.}, \bibinfo{year}{2017}.
\newblock \bibinfo{title}{Energy harvesting performance and flow structure of an oscillating hydrofoil with finite span}.
\newblock \bibinfo{journal}{Journal of Fluids and Structures} \bibinfo{volume}{70}, \bibinfo{pages}{314--326}.
\bibitem[{Kinsey and Dumas(2008)}]{kinsey2008parametric}
\bibinfo{author}{Kinsey, T.}, \bibinfo{author}{Dumas, G.}, \bibinfo{year}{2008}.
\newblock \bibinfo{title}{Parametric study of an oscillating airfoil in a power-extraction regime}.
\newblock \bibinfo{journal}{AIAA journal} \bibinfo{volume}{46}, \bibinfo{pages}{1318--1330}.
\bibitem[{Kinsey et~al.(2011)Kinsey, Dumas, Lalande, Ruel, Mehut, Viarouge, Lemay and Jean}]{kinsey2011prototype}
\bibinfo{author}{Kinsey, T.}, \bibinfo{author}{Dumas, G.}, \bibinfo{author}{Lalande, G.}, \bibinfo{author}{Ruel, J.}, \bibinfo{author}{Mehut, A.}, \bibinfo{author}{Viarouge, P.}, \bibinfo{author}{Lemay, J.}, \bibinfo{author}{Jean, Y.}, \bibinfo{year}{2011}.
\newblock \bibinfo{title}{Prototype testing of a hydrokinetic turbine based on oscillating hydrofoils}.
\newblock \bibinfo{journal}{Renewable energy} \bibinfo{volume}{36}, \bibinfo{pages}{1710--1718}.
\bibitem[{Lee et~al.(2011)Lee, Xiros and Bernitsas}]{lee2011virtual}
\bibinfo{author}{Lee, J.}, \bibinfo{author}{Xiros, N.}, \bibinfo{author}{Bernitsas, M.}, \bibinfo{year}{2011}.
\newblock \bibinfo{title}{Virtual damper--spring system for viv experiments and hydrokinetic energy conversion}.
\newblock \bibinfo{journal}{Ocean Engineering} \bibinfo{volume}{38}, \bibinfo{pages}{732--747}.
\bibitem[{Lu et~al.(2014)Lu, Xie and Zhang}]{lu2014nonsinusoidal}
\bibinfo{author}{Lu, K.}, \bibinfo{author}{Xie, Y.}, \bibinfo{author}{Zhang, D.}, \bibinfo{year}{2014}.
\newblock \bibinfo{title}{Nonsinusoidal motion effects on energy extraction performance of a flapping foil}.
\newblock \bibinfo{journal}{Renewable energy} \bibinfo{volume}{64}, \bibinfo{pages}{283--293}.
\bibitem[{Mackowski and Williamson(2017)}]{mackowski2017effect}
\bibinfo{author}{Mackowski, A.}, \bibinfo{author}{Williamson, C.}, \bibinfo{year}{2017}.
\newblock \bibinfo{title}{Effect of pivot location and passive heave on propulsion from a pitching airfoil}.
\newblock \bibinfo{journal}{Physical Review Fluids} \bibinfo{volume}{2}, \bibinfo{pages}{013101}.
\bibitem[{Mackowski and Williamson(2011)}]{mackowski2011developing}
\bibinfo{author}{Mackowski, A.W.}, \bibinfo{author}{Williamson, C.H.}, \bibinfo{year}{2011}.
\newblock \bibinfo{title}{Developing a cyber-physical fluid dynamics facility for fluid--structure interaction studies}.
\newblock \bibinfo{journal}{Journal of Fluids and Structures} \bibinfo{volume}{27}, \bibinfo{pages}{748--757}.
\bibitem[{McKinney and DeLaurier(1981)}]{mckinney1981wingmill}
\bibinfo{author}{McKinney, W.}, \bibinfo{author}{DeLaurier, J.}, \bibinfo{year}{1981}.
\newblock \bibinfo{title}{Wingmill: an oscillating-wing windmill}.
\newblock \bibinfo{journal}{Journal of energy} \bibinfo{volume}{5}, \bibinfo{pages}{109--115}.
\bibitem[{Onoue et~al.(2015)Onoue, Song, Strom and Breuer}]{onoue2015large}
\bibinfo{author}{Onoue, K.}, \bibinfo{author}{Song, A.}, \bibinfo{author}{Strom, B.}, \bibinfo{author}{Breuer, K.S.}, \bibinfo{year}{2015}.
\newblock \bibinfo{title}{Large amplitude flow-induced oscillations and energy harvesting using a cyber-physical pitching plate}.
\newblock \bibinfo{journal}{Journal of Fluids and Structures} \bibinfo{volume}{55}, \bibinfo{pages}{262--275}.
\bibitem[{Oshkai et~al.(2022)Oshkai, Iverson, Lee and Dumas}]{oshkai2022reliability}
\bibinfo{author}{Oshkai, P.}, \bibinfo{author}{Iverson, D.}, \bibinfo{author}{Lee, W.}, \bibinfo{author}{Dumas, G.}, \bibinfo{year}{2022}.
\newblock \bibinfo{title}{Reliability study of a fully-passive oscillating foil turbine operating in a periodically-perturbed inflow}.
\newblock \bibinfo{journal}{Journal of Fluids and Structures} \bibinfo{volume}{113}, \bibinfo{pages}{103630}.
\bibitem[{Platzer et~al.(2009)Platzer, Ashraf, Young and Lai}]{platzer2009development}
\bibinfo{author}{Platzer, M.}, \bibinfo{author}{Ashraf, M.}, \bibinfo{author}{Young, J.}, \bibinfo{author}{Lai, J.}, \bibinfo{year}{2009}.
\newblock \bibinfo{title}{Development of a new oscillating-wing wind and hydropower generator}, in: \bibinfo{booktitle}{47th AIAA Aerospace Sciences Meeting including the New Horizons Forum and Aerospace Exposition}, p. \bibinfo{pages}{1211}.
\bibitem[{Qadri et~al.(2019)Qadri, Shahzad, Zhao, Tang et~al.}]{qadri2019experimental}
\bibinfo{author}{Qadri, M.M.}, \bibinfo{author}{Shahzad, A.}, \bibinfo{author}{Zhao, F.}, \bibinfo{author}{Tang, H.}, et~al., \bibinfo{year}{2019}.
\newblock \bibinfo{title}{An experimental investigation of a passively flapping foil in energy harvesting mode}.
\newblock \bibinfo{journal}{Journal of Applied Fluid Mechanics} \bibinfo{volume}{12}, \bibinfo{pages}{1547--1561}.
\bibitem[{Qadri et~al.(2020)Qadri, Zhao and Tang}]{qadri2020fluid}
\bibinfo{author}{Qadri, M.M.}, \bibinfo{author}{Zhao, F.}, \bibinfo{author}{Tang, H.}, \bibinfo{year}{2020}.
\newblock \bibinfo{title}{Fluid-structure interaction of a fully passive flapping foil for flow energy extraction}.
\newblock \bibinfo{journal}{International Journal of Mechanical Sciences} \bibinfo{volume}{177}, \bibinfo{pages}{105587}.
\bibitem[{Ribeiro and Franck(2024)}]{ribeiro2024prediction}
\bibinfo{author}{Ribeiro, B.L.R.}, \bibinfo{author}{Franck, J.A.}, \bibinfo{year}{2024}.
\newblock \bibinfo{title}{Prediction of energy harvesting efficiency through a wake--foil interaction model for oscillating foil arrays}.
\newblock \bibinfo{journal}{Journal of Fluid Mechanics} \bibinfo{volume}{996}, \bibinfo{pages}{A46}.
\bibitem[{Ribeiro et~al.(2020)Ribeiro, Frank and Franck}]{ribeiro2020vortex}
\bibinfo{author}{Ribeiro, B.L.R.}, \bibinfo{author}{Frank, S.L.}, \bibinfo{author}{Franck, J.A.}, \bibinfo{year}{2020}.
\newblock \bibinfo{title}{Vortex dynamics and reynolds number effects of an oscillating hydrofoil in energy harvesting mode}.
\newblock \bibinfo{journal}{Journal of Fluids and Structures} \bibinfo{volume}{94}, \bibinfo{pages}{102888}.
\bibitem[{Ribeiro et~al.(2021)Ribeiro, Su, Guillaumin, Breuer and Franck}]{ribeiro2021wake}
\bibinfo{author}{Ribeiro, B.L.R.}, \bibinfo{author}{Su, Y.}, \bibinfo{author}{Guillaumin, Q.}, \bibinfo{author}{Breuer, K.S.}, \bibinfo{author}{Franck, J.A.}, \bibinfo{year}{2021}.
\newblock \bibinfo{title}{Wake-foil interactions and energy harvesting efficiency in tandem oscillating foils}.
\newblock \bibinfo{journal}{Physical Review Fluids} \bibinfo{volume}{6}, \bibinfo{pages}{074703}.
\bibitem[{Siala et~al.(2020)Siala, Kamrani~Fard and Liburdy}]{siala2020experimental}
\bibinfo{author}{Siala, F.F.}, \bibinfo{author}{Kamrani~Fard, K.}, \bibinfo{author}{Liburdy, J.A.}, \bibinfo{year}{2020}.
\newblock \bibinfo{title}{Experimental study of inertia-based passive flexibility of a heaving and pitching airfoil operating in the energy harvesting regime}.
\newblock \bibinfo{journal}{Physics of Fluids} \bibinfo{volume}{32}.
\bibitem[{Simeski(2017)}]{simeski2017simulations}
\bibinfo{author}{Simeski, F.}, \bibinfo{year}{2017}.
\newblock \bibinfo{title}{Simulations of Hydrofoil Arrays with Applications in Energy Harvesting}.
\newblock Ph.D. thesis. Bachelor’s thesis, Brown University.
\bibitem[{Simpson et~al.(2008)Simpson, Hover and Triantafyllou}]{simpson2008experiments}
\bibinfo{author}{Simpson, B.J.}, \bibinfo{author}{Hover, F.S.}, \bibinfo{author}{Triantafyllou, M.S.}, \bibinfo{year}{2008}.
\newblock \bibinfo{title}{Experiments in direct energy extraction through flapping foils}, in: \bibinfo{booktitle}{ISOPE International Ocean and Polar Engineering Conference}, \bibinfo{organization}{ISOPE}. pp. \bibinfo{pages}{ISOPE--I}.
\bibitem[{Sorensen(2024)}]{sorensen2024developing}
\bibinfo{author}{Sorensen, A.}, \bibinfo{year}{2024}.
\newblock \bibinfo{title}{Developing a four-axis cyber-physical traverse system for highly dynamic experimental fluid mechanics studies}.
\newblock Master's thesis. Iowa State University.
\bibitem[{Su and Breuer(2019)}]{su2019resonant}
\bibinfo{author}{Su, Y.}, \bibinfo{author}{Breuer, K.}, \bibinfo{year}{2019}.
\newblock \bibinfo{title}{Resonant response and optimal energy harvesting of an elastically mounted pitching and heaving hydrofoil}.
\newblock \bibinfo{journal}{Physical Review Fluids} \bibinfo{volume}{4}, \bibinfo{pages}{064701}.
\bibitem[{Su et~al.(2019)Su, Miller, Mandre and Breuer}]{su2019confinement}
\bibinfo{author}{Su, Y.}, \bibinfo{author}{Miller, M.}, \bibinfo{author}{Mandre, S.}, \bibinfo{author}{Breuer, K.}, \bibinfo{year}{2019}.
\newblock \bibinfo{title}{Confinement effects on energy harvesting by a heaving and pitching hydrofoil}.
\newblock \bibinfo{journal}{Journal of Fluids and Structures} \bibinfo{volume}{84}, \bibinfo{pages}{233--242}.
\bibitem[{Teng et~al.(2016)Teng, Deng, Pan and Shao}]{teng2016effects}
\bibinfo{author}{Teng, L.}, \bibinfo{author}{Deng, J.}, \bibinfo{author}{Pan, D.}, \bibinfo{author}{Shao, X.}, \bibinfo{year}{2016}.
\newblock \bibinfo{title}{Effects of non-sinusoidal pitching motion on energy extraction performance of a semi-active flapping foil}.
\newblock \bibinfo{journal}{Renewable energy} \bibinfo{volume}{85}, \bibinfo{pages}{810--818}.
\bibitem[{Veilleux and Dumas(2017)}]{veilleux2017numerical}
\bibinfo{author}{Veilleux, J.C.}, \bibinfo{author}{Dumas, G.}, \bibinfo{year}{2017}.
\newblock \bibinfo{title}{Numerical optimization of a fully-passive flapping-airfoil turbine}.
\newblock \bibinfo{journal}{Journal of Fluids and Structures} \bibinfo{volume}{70}, \bibinfo{pages}{102--130}.
\bibitem[{Wang et~al.(2021)Wang, Du, Zhao, Thompson and Sun}]{wang2021pivot}
\bibinfo{author}{Wang, Z.}, \bibinfo{author}{Du, L.}, \bibinfo{author}{Zhao, J.}, \bibinfo{author}{Thompson, M.C.}, \bibinfo{author}{Sun, X.}, \bibinfo{year}{2021}.
\newblock \bibinfo{title}{Pivot location and mass ratio effects on flow-induced vibration of a fully passive flapping foil}.
\newblock \bibinfo{journal}{Journal of Fluids and Structures} \bibinfo{volume}{100}, \bibinfo{pages}{103170}.
\bibitem[{Xiao et~al.(2012)Xiao, Liao, Yang and Peng}]{xiao2012motion}
\bibinfo{author}{Xiao, Q.}, \bibinfo{author}{Liao, W.}, \bibinfo{author}{Yang, S.}, \bibinfo{author}{Peng, Y.}, \bibinfo{year}{2012}.
\newblock \bibinfo{title}{How motion trajectory affects energy extraction performance of a biomimic energy generator with an oscillating foil?}
\newblock \bibinfo{journal}{Renewable energy} \bibinfo{volume}{37}, \bibinfo{pages}{61--75}.
\bibitem[{Xiao and Zhu(2014)}]{xiao2014review}
\bibinfo{author}{Xiao, Q.}, \bibinfo{author}{Zhu, Q.}, \bibinfo{year}{2014}.
\newblock \bibinfo{title}{A review on flow energy harvesters based on flapping foils}.
\newblock \bibinfo{journal}{Journal of fluids and structures} \bibinfo{volume}{46}, \bibinfo{pages}{174--191}.
\bibitem[{Young et~al.(2013)Young, Ashraf, Lai and Platzer}]{young2013numerical}
\bibinfo{author}{Young, J.}, \bibinfo{author}{Ashraf, M.A.}, \bibinfo{author}{Lai, J.C.}, \bibinfo{author}{Platzer, M.F.}, \bibinfo{year}{2013}.
\newblock \bibinfo{title}{Numerical simulation of fully passive flapping foil power generation}.
\newblock \bibinfo{journal}{AIAA journal} \bibinfo{volume}{51}, \bibinfo{pages}{2727--2739}.
\bibitem[{Zhao et~al.(2021)Zhao, Qadri, Wang and Tang}]{zhao2021flow}
\bibinfo{author}{Zhao, F.}, \bibinfo{author}{Qadri, M.M.}, \bibinfo{author}{Wang, Z.}, \bibinfo{author}{Tang, H.}, \bibinfo{year}{2021}.
\newblock \bibinfo{title}{Flow-energy harvesting using a fully passive flapping foil: A guideline on design and operation}.
\newblock \bibinfo{journal}{International Journal of Mechanical Sciences} \bibinfo{volume}{197}, \bibinfo{pages}{106323}.
\bibitem[{Zhu et~al.(2021a)Zhu, Zhu and Zhang}]{zhu2021improve}
\bibinfo{author}{Zhu, J.}, \bibinfo{author}{Zhu, M.}, \bibinfo{author}{Zhang, T.}, \bibinfo{year}{2021}a.
\newblock \bibinfo{title}{Improve the performance of a semi-active flapping airfoil power generator by adjusting both offsetting mass center displacement and changing pitching axis position}.
\newblock \bibinfo{journal}{Energy Reports} \bibinfo{volume}{7}, \bibinfo{pages}{5074--5085}.
\bibitem[{Zhu(2011)}]{zhu2011optimal}
\bibinfo{author}{Zhu, Q.}, \bibinfo{year}{2011}.
\newblock \bibinfo{title}{Optimal frequency for flow energy harvesting of a flapping foil}.
\newblock \bibinfo{journal}{Journal of fluid mechanics} \bibinfo{volume}{675}, \bibinfo{pages}{495--517}.
\bibitem[{Zhu et~al.(2021b)Zhu, Mathai and Breuer}]{zhu2021nonlinear}
\bibinfo{author}{Zhu, Y.}, \bibinfo{author}{Mathai, V.}, \bibinfo{author}{Breuer, K.}, \bibinfo{year}{2021}b.
\newblock \bibinfo{title}{Nonlinear fluid damping of elastically mounted pitching wings in quiescent water}.
\newblock \bibinfo{journal}{Journal of Fluid Mechanics} \bibinfo{volume}{923}, \bibinfo{pages}{R2}.
\bibitem[{Zhu et~al.(2020)Zhu, Su and Breuer}]{zhu2020nonlinear}
\bibinfo{author}{Zhu, Y.}, \bibinfo{author}{Su, Y.}, \bibinfo{author}{Breuer, K.}, \bibinfo{year}{2020}.
\newblock \bibinfo{title}{Nonlinear flow-induced instability of an elastically mounted pitching wing}.
\newblock \bibinfo{journal}{Journal of Fluid Mechanics} \bibinfo{volume}{899}, \bibinfo{pages}{A35}.

\end{thebibliography}
\end{document}